\documentstyle[psfig, rotate]{l-aa}
\begin{document}
\title{The optical spectrum of HR~4049
\thanks{Based on observations obtained with
the WHT/UES (La Palma) and CAT/CES (ESO)}
\thanks{Fig.~9 and Table~11 are only available in electronic form
at the CDS via anonymous ftp 130.79.128.5}}
\subtitle{(includes a line identification from 3650 to 10850~\AA)}
\author{Eric J. Bakker\inst{1,2} \and
        Frank L.A. Van der Wolf\inst{1,2}\and
        Henny J.G.L.M. Lamers\inst{1,2}\and
        Austin F. Gulliver\inst{3}\and
        Roger Ferlet\inst{4}\and
        Alfred Vidal-Madjar\inst{4}}
\offprints{Eric J. Bakker, present address:
        Astronomy Department,
        University of Texas,
        Austin, TX 78712-1083,
        U.S.A., ebakker@astro.as.utexas.edu}
\institute{Astronomical Institute,
           University of Utrecht,
           P.O.box 80000,
           NL-3508 TA Utrecht,
           The Netherlands
           \and
           SRON Space Research Laboratory,
           Sorbonnelaan 2, NL-3584 CA
           Utrecht,
           The Netherlands
           \and
           Department of Physics \&\ Astronomy,
           Brandon University,
           Brandon,
           MB R7A 6A9,
           Canada
           \and
           Institut d'Astrophysique de Paris,
           CNR 98 bis Boulevard Arago,
           F-75014 Paris,
           France}
\date{received: February 1 1995, accepted: July 25 1995}
\thesaurus{08(02.12.2;             
              02.12.3;             
              08.02.1;             
              08.03.2;             
              08.09.2: HR~4049;    
              08.16.4)}            

\maketitle

\begin{abstract}
High-resolution optical spectra  (UES/WHT)
of the extreme metal-poor post-AGB  star HR~4049 were obtained at
four different orbital phases. The spectra cover the wavelength
region from 3650~\AA~ to 10850~\AA~ at a resolution of $R=5.2 \times 10^{4}$.
These observations are supplemented with four high-resolution spectra
of the NaI D1 \& D2 and CaII K lines at $R \approx 10^{5}$ (CAT/CES).
The optical spectrum shows  217 spectral lines:
the Balmer series (H$\alpha$ - H35), the Paschen series (P9 - P23), NI,
OI and numerous CI lines.
We show that the lines of H$\alpha$, H$\beta$, H$\gamma$ and NaI D
show significant changes in profile between different observation
dates. Nine components were identified in the profile of the NaI D lines
of which three are circumstellar and six interstellar.
The stronger CI lines are asymmetric and we derive a post-AGB mass-loss
of
\.{M}$= 6\pm 4 \times 10^{-7}$~M$_{\odot}$~yr$^{-1}$ from the asymmetry.
The [OI] 6300~\AA~ line
has been detected in emission
at the system velocity and we argue that the emission is
from an almost edge-on disk with a radius of about $20~R_{\ast}$.
\end{abstract}

\section{Introduction}

HR~4049 (HD~89353) (see Table~\ref{art7tab-hr4049})
is a member of the family of extremely metal poor post-AGB stars
in a close binary system with a yet unseen companion star
(Waelkens {\it et al.} \cite{art7waelkenslamers},
Van Winckel {\it et al. } \cite{art7winckel}).
The semi-major axis of the system of
$a \sin i = 3 R_{\ast}$ (Van Winckel {\it et al. } \cite{art7winckel})
is smaller than the radius of HR~4049 during the preceding
AGB phase ($ \approx 200 R_{\ast}$).
This means that there was a common envelope evolution without
spiraling in of the secondary. The time scale for circularization in a
common envelope phase is very short. The fact that HR~4049 has
an eccentricity of $e=0.31$ suggest that binary interaction, possible
with a circumsystem disk, are able to de-circularize the orbit.
With a semi-major axis of the binary smaller than
the size of the star during the previous AGB phase it
is not clear how this system could have survived the AGB.

The abundance pattern with iron a factor $10^{-4.8}$ under abundant
and C, N, O and S close to solar, cannot be explained in terms
of nucleo-synthesis products, but resembles that of the depleted
interstellar gas. It is thought that the low abundance of the metals is
due to a mechanism by which first the metals are condensed on dust in the
circumstellar environment, followed by a separation of the dust and gas.
The metal-poor gas then falls on the star and forms a
photosphere of low metalicity, containing at least $10^{-5}$~M$_{\odot}$,
which can only be
sustained as long as there is no significant convection or strong
stellar wind (Mathis \& Lamers \cite{art7mathislamers};
Waters {\it et al. } \cite{art7waterstrams}).
Taking a typical post-AGB mass-loss rate of
\.{M}$\approx 10 ^{-7}$ to $10^{-8}$~M$_{\odot}$~yr$^{-1}$, the depleted
photosphere would disperse in about a 100~years.

With this study we make a first attempt to study variations
of the complete optical spectrum.  The basis for this
study is the line identification list presented in
Table~11 (published at CDS), in which the profile parameters and
the identification are given for all observed spectral features
from 3650~\AA~ to 10850~\AA~. The spectrum with the identifications
is shown in Fig.~9 (at CDS).

\begin{table}
\caption{HR~4049 (HD~89353; SAO~178644)}
\label{art7tab-hr4049}
\centerline{\begin{tabular}{lll}
\hline
                &                             &reference \\
\hline
                &                             &    \\
$m_{v}$         &+5.520                       &1   \\
epoch           &JD($\phi=0.0$)=$2447235\pm3$ &1   \\
period $\Pi$    &$429\pm2$~d.                 &2   \\
$v_{\rm system}$&$-32.9\pm0.7$~km~s$^{-1}$    &2   \\
$e$             &0.31                         &2   \\
$a \sin i$      &0.583~AU=$3 R_{\ast}$        &2   \\
$f(m)$          &0.143~M$_{\odot}$            &2   \\
$T_{\rm eff}$   &7500~K                       &3   \\
$\log g$        &1.0~cm~s$^{-2}$              &3   \\
$M_{\ast}$      &0.54~M$_{\odot}$             &4   \\
$R_{\ast}$      &38~R$_{\odot}$               &4   \\
$L_{\ast}$      &$4.1\times10^{3}$~L$_{\odot}$&    \\
                &                             &    \\
\hline
\end{tabular}}
\begin{centering}
1: Waelkens {\it et al. } \cite{art7waelkenslamers};
2: van Winckel {\it et al. } \cite{art7winckel};
3: Lambert {\it et al. }  \cite{art7lamberthinkle};
4: Trams           \cite{art7trams}
\end{centering}
\end{table}

The observations and data reduction are discussed in
Sect.~\ref{art7sec-obs}.
In Sect.~\ref{art7sec-meth}
a description is given of the method of line identification
and the criteria used for identification. The resulting list is
given in Table~11 (at CDS) and Fig.~9 (at CDS).
In Sect.~\ref{art7sec-sel}
we discuss a number of
selected lines: CI, NI and OI,
Hydrogen lines,  resonance lines of NaI and CaII and
the [OI] at 6300~\AA~. In Sect.~\ref{art7sec-dis} the post-AGB mass-loss
rate is derived from the asymmetry of the CI line profile and
in Sect.~\ref{art7sec-con} the conclusions of this work are summarized.

\section{The observations and data reduction}
\label{art7sec-obs}

\subsection{WHT/UES observations}

HR~4049 was observed from La Palma with the 4.2m William Herschel Telescope
(WHT)
and the Utrecht Echelle Spectrograph (UES).
A description of the WHT/UES is given by Unger (\cite{art7unger}).
The first observation was made during
commissioning time in February 1992. This observation gave rise to an
observation campaign from February to June 1993 in service time.
A log of the WHT/UES observations is given in Table~\ref{art7tab-obswht}
where phases are calculated using the orbital parameters of
Table~\ref{art7tab-hr4049}.
All observations with the WHT/UES were obtained with echelle 31
(31.6 grooves per mm), except on
February 11 1993 when we used echelle 79.
The wavelength coverage of echelle 31 is roughly a  factor 2.5
larger than that of echelle 79.

\begin{table*}
\caption{Log of the WHT/UES observations of HR~4049}
\label{art7tab-obswht}
\centerline{\begin{tabular}{lllll}
\hline
                    &Feb. 25 92
                    &Feb. 11 93
                    &Mar.  8 93
                    &Apr.  5 93 \\
\hline
                    &          &          &          &          \\
JD                  & 2448678  & 2449030  & 2449055  &2449083   \\
$\phi$              &   0.36   &    0.18  &    0.24  &  0.31    \\
$\delta v_{\oplus}$ &9.4~km~s$^{-1}$&14.4~km~s$^{-1}$&4.4~km~s$^{-1}$
&-7.7~km~s$^{-1}$ \\
echelle             &   31.6   &    79    &    31.6  &    31.6  \\
CCD                 &   EEV3   &  TEK1    &    EEV6  &    TEK1  \\
flatfields         &   No     &   Yes    &     Yes  &    Yes   \\
                    &          &          &          &          \\
\hline
$\lambda_{c}$       &run (code)&run (code)&run (code)&run (code)\\
\hline
                    &          &          &          &          \\
7128.9~\AA          &41655(1.1)&72597(2.1)&74256(3.1)&75240(4.1)\\
                    &90 sec.   &90 sec.   &90 sec.   &120 sec.  \\
5260.9~\AA          &41656(1.2)&72594(2.2)&74262(3.2)&75238(4.2)\\
                    &90 sec.   &120 sec.  &90 sec.   &120 sec.  \\
4020.0~\AA          &41660(1.3)&72592(2.3)&74269(3.3)&75230(4.3)\\
                    & 300 sec. &60 sec.   &1500 sec. &120 sec.  \\
                    &          &          &          &          \\
\hline
\end{tabular}}
\centerline{$v_{\odot} = v_{\rm obs} + \delta v_{\oplus}$}
\end{table*}

Three central wavelength settings were used: 4020~\AA, 5260~\AA~ and
7127~\AA~ covering a wavelength range from 3650~\AA~ to 10850~\AA.
The wavelength coverage of echelle 31 are 5350-10850~\AA~ for
$\lambda_{c}=7129$~\AA, 4380-6965~\AA~ for $\lambda_{c}=5261$~\AA~
and $3590-4665$~\AA~ for $\lambda_{c}=4020$~\AA.
There is overlap of spectral orders in the
blue, but not in the red ($\lambda \geq 7000$~\AA).
A complete spectrum
could be obtained within no more than 40 minutes at
a resolution of $R=5.2 \times 10^{4}$ corresponding to 6~km~s$^{-1}$.
The spectra were calibrated using
a Th-Ar lamp. The internal accuracy of the wavelength scale
is about  6~m\AA.
After extraction of the echellograms, the spectral data has an external
accuracy  of about 6~km~s$^{-1}$.
For $\lambda \geq 9500$~\AA~ no wavelength calibration of the Th-Ar lamp
was available at the time of reduction, which give rise to a small
wavelength drift up to about 0.2~\AA~ ($\approx 6$~km~s$^{-1}$).

Some problems were encountered in flatfielding the UES spectra.
Neutral density filters caused fringe patterns on the flatfields.
These filters were applied to diminish the intensity of the
Tungsten lamp, used for the flatfield images.
The TEK1 CCD ($1024 \times 1024$ pixels) produces a fringe pattern
caused by its coating. This fringing starts to become important
for $\lambda \geq 6000$~\AA.
{}From 6500~\AA~ to 9000~\AA~ the fringe amplitude is about 2 to 3~\%.
{}From 9000~\AA~ to
10000~\AA~ it becomes dramatically worse, going up to 10 and even 20~\%
(Tinbergen \cite{art7tinbergen}). This fringing
pattern occurs in both the target image and flatfield so that it can
be removed.

We checked the wavelength
resolution of the UES by looking at unresolved telluric absorption lines
at 6880~\AA~ and
found for the rotation lines of
telluric O$_{2}$ absorption a full-width-full-maximum of
$\approx 6$~km~s$^{-1}$, which is equivalent to a resolution of
$R = 5 \times 10^{4}$.
Using the identification and wavelength in the solar spectrum atlas
from Moore {\it et al. }
(\cite{art7mooreminnaert}) of the O$_{2}$ telluric band
we find a small shift of $0.1\pm0.3$~km~s$^{-1}$
with a standard deviation of $\sigma=2.5$~km~s$^{-1}$.

In this study we have concentrated on the the observations from
March 8 1993.
The observations in February and April 1993 gave identical identifications.
Differences occured because {\it i)} different detectors were used; {\it ii)}
different echelles were used giving rise to a slightly
different wavelength coverage; {\it iii)}
variations in the S/N ratio occured between different observations.

The binary motion of HR~4049 causes a velocity shift for photospheric and wind
lines, but not for interstellar and circumstellar lines.

\subsection{CAT/CES observations}

HR~4049 was observed at four different dates at high-resolution
and high signal-to-noise
ratio with the 1.4m Coud\'{e} Auxiliary Telescope
(CAT) at ESO. The CAT was equipped with a Coud\'{e} Echelle
Spectrometer (CES) and a Reticon detector giving a spectral
resolving power of $R=10^{5}$
which corresponds to 3~km~s$^{-1}$. The wavelength
calibration is achieved by means of many emission lines of
Th-Ar hollow cathode lamp, which yields and internal accuracy
better than 0.3~km~s$^{-1}$ and an external accuracy of about 1~km~s$^{-1}$
(Ferlet \& Dennefeld \cite{art7ferden}).

The CAT/CES spectra with a typical wavelength coverage
of 30~\AA~ were centered around
the resonance NaI D1~\&~D2 and CaII~K lines
and  have a factor two higher resolution than the WHT/UES.
The spectra centered around the NaI  lines have been divided by
the spectrum of a bright template star in order to remove the
numerous atmospheric H$_{2}$O lines in this spectral range.
A log of the CAT/CES observations is given in Table~\ref{art7tab-obscat}.

\begin{table}
\caption{Log of the CAT/CES observations of HR~4049}
\label{art7tab-obscat}
\centerline{\begin{tabular}{llllll}
\hline
Date      & JD    &$\delta v_{\oplus}$&Int.&$\phi$ & remark    \\
          &       &[km/s]             &[s] &       &           \\
\hline
          &       &                   &    &    &       \\
Jan. 11 86&2446442&20.8               &3600&0.15& CaII K\\
Jan. 12 86&2446443&20.4               &1800&0.15& NaI D \\
Nov. 20 86&2446755&22.9               &2200&0.88& CaII K\\
Nov. 27 86&2446762&23.8               &3000&0.90& NaI D \\
Feb. 23 87&2446850& 6.2               &1800&0.10& NaI D \\
          &       &                   &    &    &       \\
\hline
\end{tabular}}
\end{table}

\section{The line identification}
\label{art7sec-meth}

\subsection{The method}

The first step in the line identification process
was checking those
ions expected to be present in the photosphere of a metal poor
supergiant with $T_{\rm eff} = 7500$~K.
This showed that the main contribution of the
absorption lines are from HI, CI, NI, OI and
the resonance lines of NaI and CaII.
The second step in the line identification process was the use of this
line identification list as input to the interactive line identification
programs IDPORC and LINEID. A description of these
programs is given by Gulliver and Stadel (\cite{art7gulliverstadel}).
In both steps the same selection criteria for identification were used.

In the first step the identification was made by hand using the multiplet
table of Moore (\cite{art7moore}).
A multiplet was considered detected
if the lines within one multiplet were observed with relative equivalent
width in acceptable agreement
with the intensity values as given in the multiplet tables. A constraint
on the identification is that the
Doppler velocities of lines within one multiplet should be
the same. Most of the absorption lines were identified
using these two criteria. The remaining features, mostly single lines, were
identified by comparing the
central absorption wavelengths with several finding lists.

This first step left a great number of unidentified weak
lines. In the second step
LINEID and IDPROC were used to check the already identified
lines and to give
possible identifications of the unidentified lines. The identification program
uses an extensive spectral database. After running the program, the total
number
of unidentified lines was reduced by a factor 10! Most of the previously
unidentified lines appeared to be due to CI multiplets, which were not
described in the multiplet tables that were used in the first step.

The identification list of absorption lines of the optical spectrum of
HR~4049 is given in Table~11 (at CDS) and overplotted in Fig.~9 (at CDS).
The central wavelengths of the  absorption lines
were measured by fitting Gaussian profiles to the observed profiles.
Other parameters like central depth ($D$),
Full-Width-Half-Maximum ($FWHM$) and
equivalent width ($W_{\lambda}$) were derived from these fits.
In case of weak or asymmetric lines an eyeball fit was made to determine the
central
wavelength and central depth. The equivalent width was measured by
integrating over the whole line profile.

In fitting the lines with Gaussian profiles we implicitly assumed that
the line broadening is mainly due to Doppler broadening.
For CI, NI and OI lines this seems to be correct, but the method fails in the
case of Hydrogen lines. The line wings are formed at different velocities with
respect to the line cores, which makes the profiles asymmetric. In these
cases the equivalent width has been calculated by integration over the whole
line profile.

If an absorption feature was blended by other features, an estimate of
the central
wavelength and other parameters was made by fitting several Gaussian curves
to the observed profile.
A fit to a blended profiles is not always unique and
the values of $D$, $FWHM$ and
$W_{\lambda}$ are less accurate than those for
isolated lines.

\begin{table*}
\caption{C~I lines in the spectrum of HR~4049 compared with
observations of Waelkens {\it et al. } (1991b) in HD~52961 (see text)}
\label{art7tab-hd52961}
\centerline{\begin{tabular}{cc|cc|cc|c}
\hline
&&\multicolumn{2}{|c|}{HR~4049} & \multicolumn{2}{c|}{HD~52961}&\\
$\lambda_{lab}$& Mult. & $\lambda_{\rm obs}$ &$W_{\lambda}$
& $\lambda_{\rm obs}$
&$W_{\lambda}$& $|v_{dif}|$ \\
{}~[\AA] & & [\AA] &  [m\AA] & [\AA] &  [m\AA] &  [km~s$^{-1}$]\\
\hline
         &       &          &    &          &    &     \\
4212.342 & 18.12 & 4212.057 & 11 & 4212.487 & 35 & 30.6\\
4223.360 & 18.11 & 4222.934 & 16 & 4223.383 & 56 & 31.9\\
4466.476 & 18.07 & 4466.138 & 15 & 4466.616 & 47 & 32.1\\
4477.472 & 18.06 & 4477.149 & 15 & 4477.635 & 46 & 32.6\\
4478.588 & 18.06 & 4478.273 & 30 & 4478.840 & 96 & 38.0\\
         &       &          &    &          &    &     \\
\hline
\end{tabular}}
\end{table*}

Waelkens {\it et al. } (\cite{art7waelkenswinckel}) found several absorption
lines in the
spectrum of HD~52961 which they could not identify.
The extreme metal poor star HD~52961 belongs to the same group of
post-AGB stars as HR~4049. It is therefore interesting to note
that the unidentified lines in HD~52961 are also present in the
spectrum of HR~4049 and identified as CI.
The equivalent widths of
these lines in HR~4049 are roughly a factor three lower than in HD~52961
(Table~\ref{art7tab-hd52961}).

Waelkens {\it et al. } (\cite{art7waelkenswinckel})
suggest molecular lines of CH$^{+}$
in the wavelength ranges of 4204-4238~\AA~
and 4284-4318~\AA.  From the work by Bakker {\it et al. }
(\cite{art7bakkeredin})
we know that two other post-AGB stars, HD~44179 (Red Rectangle) and
HD~213985, show the $\rm A^{1}\Pi-X^{1}\Sigma^{+}$ (0,0) band
of CH$^{+}$ at 4240~\AA in emission and absorption respectively.
There are at least ten other post-AGB stars which
show C$_{2}$ and CN absorption by gas in the AGB remnant
(Bakker {\it et al. } \cite{art7bakkeredin}).
We checked this wavelength region carefully for
the presence of possible molecular absorption or emission lines and
could not find any. This non-detection of molecules might
be important as it distinguishes HR~4049 form other post-AGB stars.

We confirm the detection of the HeI 4471~\AA~ absorption lines
(Waelkens {\it et al. } \cite{art7waelkenswinckel}) and present the first
detection of
the forbidden [OI] line at 6300.311~\AA~ (see Fig.~\ref{art7fig-oi}).
Lambert {\it et al. }
(\cite{art7lamberthinkle}) referred to this line in HR~4049
but they were not able to detect it. The  UES spectrum have a S/N-ratio
$\approx 100$
and clearly shows an emission line at the system velocity of HR~4049.

A normal B9 supergiant shows no NaI D1 \& D2 lines and a very small CaII K
photospheric lines. All three lines are dominant in the optical spectrum of
HR~4049 and are thus from non-photospheric origin.

\subsection{Line identifications list}
\label{art7subsec-tablist}

\noindent
{\sl Description of line identification list}

The format of the Table~11 (at CDS) is listed below.
The parameters of the atomic and ionic transitions are
listed in column 1 through 4. They were all taken from the input database
to the LINEID program as described in Gulliver and Stadel
(\cite{art7gulliverstadel}).

\begin{enumerate}
\item $\lambda_{\rm lab}$ [\AA]: the laboratory wavelength of the
identified absorption  line. An $\ast$ is given when the runnumber
is not 3 (March 8 93).

\item multiplet: the element and multiplet number of the given absorption line.

\item $\chi$ [eV]: the lower level excitation energy of the transition.
If an accurate value  could not be found in the literature,
the entry is left blank.

\item $\log gf$: the oscillator strength of the transition.
If an accurate value  could not be found in the literature,
the entry is left blank.

\item $\lambda_{\rm obs}$ [\AA]: the observed central wavelength of the
absorption
feature. The estimated accuracy is about 3~km~s$^{-1}$,
based on the S/N ratio and internal
accuracy of the calibration method.

\item $D$ [\%]: the observed depth of the central absorption relative
to the continuum. A line to zero intensity gives a depression of
100\%. A mean error of 2\%\ is adopted.

\item $W_\lambda$ [m\AA]: the observed equivalent width.
The estimated error on the equivalent width is
10\%. If one of the wings of
the profile could not be measured (due to blending or no wavelength
coverage of the spectrum) this entry is left blank.

\item $FWHM$ [km~s$^{-1}$]: Full-Width-Half-Maximum of the absorption feature.
The accuracy
is dependent on the S/N ratio and is estimated at 3~km~s$^{-1}$.

\item $v_{rad}$ [km~s$^{-1}$]: radial heliocentric velocity for
the identified absorption
feature. The accuracy is 3~km~s$^{-1}$.

\item run: the number of the spectra are give in Table~\ref{art7tab-obswht}.
They refer to the number which was assigned to this spectrum during the
observations.

\item $Q$: quality factor for the identification of the absorption feature.
Lines
marked with {\bf 3} are positively uniquely identified.
A quality of {\bf 2}  means that line is lightly blended.
Quality {\bf 1} lines are severely polluted or very weak and can not be used
for further studies.

\item remarks on the line identification.
{\bf a}: asymmetric line profile;
{\bf b}: line profile is blended;
{\bf f}: forbidden line;
{\bf IS}: interstellar;
{\bf CS}: circumstellar
\end{enumerate}

The number of lines identified for the species present in the
spectrum of HR~4049 are summarized in Table~\ref{art7tab-stat}.

\begin{table}
\caption{Overview of observed spectral lines in the
optical spectrum of HR~4049}
\label{art7tab-stat}
\centerline{\begin{tabular}{lrl}
\hline
ion & no. of & remark \\
    & lines  &        \\
\hline
        &   &                                 \\
{}~H~I    &48 &H$\alpha$ - H35 and  P9 -P23     \\
{}~HeI    & 1 &4471~\AA                         \\
{}~C~I    &82 &                                 \\
{}~N~I    &14 &                                 \\
{}~O~I    &20 &                                 \\
{}~[OI]   & 1 &6300~\AA                         \\
{}~NaI    &14 &7 comp. in D                     \\
{}~S~I    & 9 &                                 \\
{}~CaII   &12 &7 comp. in K \& 3 in H \& CaII(2)\\
{}~unident&16 &                                 \\
{}~       &   &                                 \\
{}~Total  &217&                                 \\
        &   &                                 \\
\hline
\end{tabular}}
\end{table}

\section{Selected line profiles}
\label{art7sec-sel}

\subsection{CI, NI and OI lines profiles}
\label{art7subsec-cno}

\begin{figure*}
\centerline{\hbox{\psfig{figure=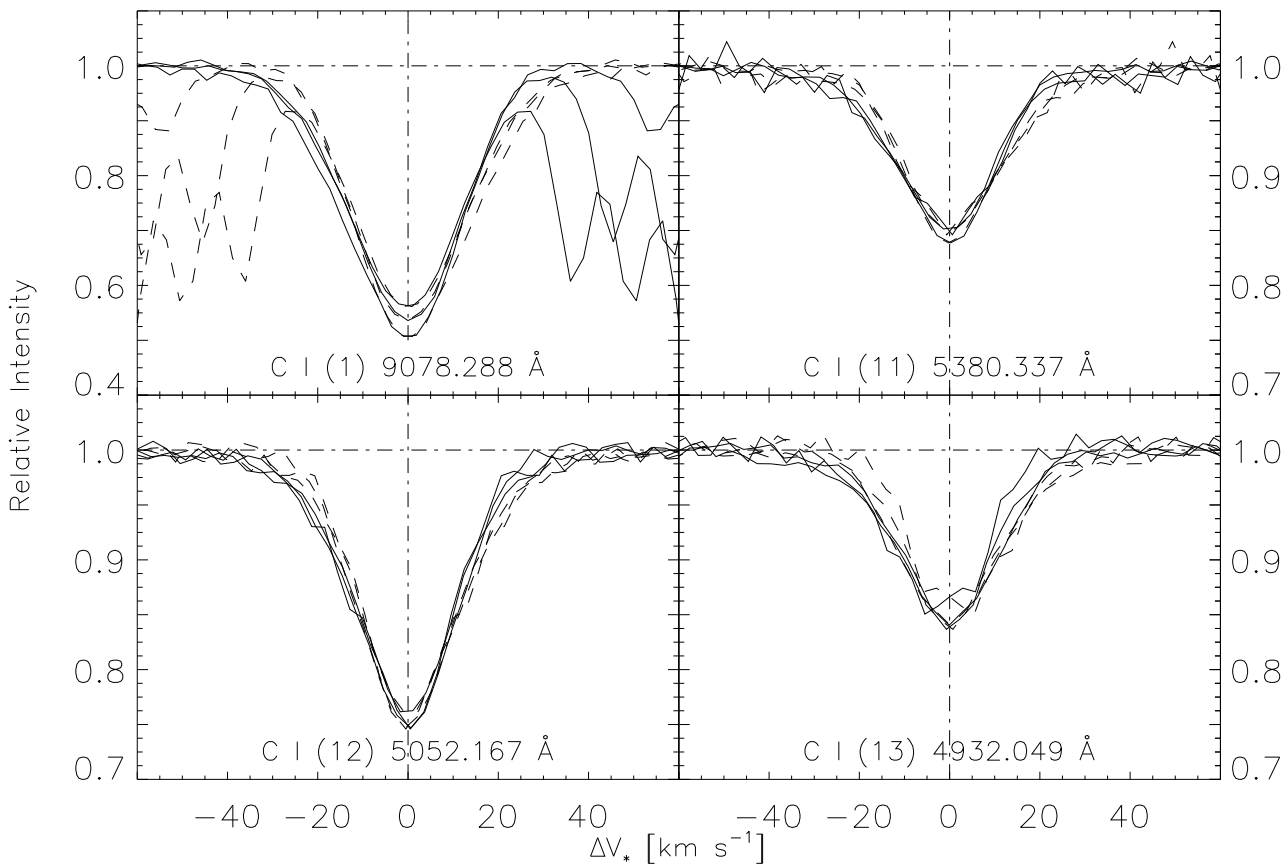,width=\textwidth}}}
\caption{Asymmetric CI lines. The dashed profiles represent the
mirrored line profiles, to show the asymmetry. These mirrored profiles
show that the absorption
is stronger at the blue side of the profile at velocity
$\Delta v_{\ast} \simeq -10$ to -40~km~s$^{-1}$}
\label{art7fig-clines}
\end{figure*}

The majority (54~\%) of absorption lines in the optical spectrum of HR~4049
are from neutral C, N and O. These lines were used
to make an accurate determination of the stellar velocity.
This assumes that the CI, NI and OI
lines do not experience any emission as the Hydrogen
lines do. To check whether this is a reasonable assumption we have
selected four strong CI lines and plotted their lines profiles at
three different phases during one orbital cycle
(solid profiles in Fig.~\ref{art7fig-clines}).

Only the CI line at $\lambda = 9078.288$~\AA~  shows a change in strength
whereas the other three CI lines do not show any
change in strength. As the CI lines at $\lambda = 9078.288$~\AA~  is in
a part of the spectrum with numerous telluric lines, we attribute this
change in intensity to a small error in the adopted continuum level.
We conclude that the profiles
did not vary in depth and equivalent width
during the period we observed HR~4049.

To see whether the CI lines have a wind contribution we have mirrored
the lines profiles on the average CI, NI and OI velocity (stellar velocity)
and over plotted them (dashed profiles) with the observed lines profiles.
We clearly see that in all twelve profiles the mirrored profile shows less
absorption at negative velocities (-10 to -40~km~s$^{-1}$)
than the observed profile.
In other words: the CI lines have stronger blue wings then red wings.
This suggests a small contribution of the wind to the CI lines.
We have also studied the N and O line profiles and
found similar results. In order to exclude that the asymmetries
are of instrumental origin, we have looked at the same
lines in the spectrum of $\eta$ Auri (A8Ia) and found that in this
spectrum all (unblended) lines are symmetric. The symmetry is therefore
not instrumental.

{}From Fig.~\ref{art7fig-clines} we conclude that the
CI lines are a asymmetric, but the asymmetry does not seem to
have changed during the period we observed HR~4049.
It would be very interesting to study the asymmetry of the CI,
NI and OI lines as function of orbital phase.

\begin{figure}
\centerline{\hbox{\psfig{figure=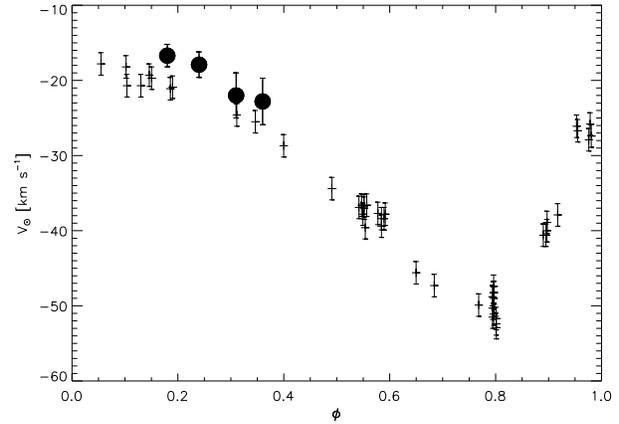,width=\columnwidth}}}
\caption{Radial velocity curve for HR~4049.
The  filled circles represent data obtained in this study}
\label{art7fig-radvel}
\end{figure}

In Table~\ref{art7tab-cno}
we have summarized the velocities of the CI, NI and OI lines.
The asymmetry is too small to be of any importance for the determination
of the stellar velocity from the CI, NI and OI lines.
As expected there
is no difference found between the average velocities of CI, NI and OI
lines. The last column gives the adopted stellar velocity.
Fig.~\ref{art7fig-radvel} shows the stellar velocities versus orbital phase
as determined from the work of Van Winckel {\it et al. } (\cite{art7winckel}).
The new data points (solid dots) fit the overall radial velocity curve
very well.

\begin{table*}
\caption{Radial  velocities of the CI, NI and OI lines
at different orbital phases
with the standard deviation. Column 6 gives the averaged photospheric velocity
based on the CI, NI and OI lines}
\label{art7tab-cno}
\centerline{\begin{tabular}{lccccc}
\hline
Date & $\phi$ & C~I    &  N~I   & O~I    & $v_{\rm CI,NI,OI}$ \\
\cline{3-6}
     &        & \multicolumn{4}{c}{[km~s$^{-1}$]}             \\
\hline
             &    &            &                &            &             \\
Febr. 12 1993&0.18&$-16.1\pm2.5$&$-15.3\pm0.1^1$&$-17.4\pm2.4$&$-16.7\pm2.5$\\
Mar.   8 1993&0.24&$-17.5\pm1.9$&$-18.3\pm2.0  $&$-18.0\pm1.1$&$-17.9\pm1.7$\\
Apr.   6 1993&0.31&$-20.8\pm2.6$&$-24.3\pm3.8  $&$-21.0\pm2.6$&$-22.0\pm3.0$\\
Febr. 26 1992&0.36&$-23.0\pm4.0$&$-23.6\pm2.7  $&$-23.6\pm3.1$&$-22.8\pm4.7$\\
             &    &            &                &            &             \\
\hline
\end{tabular}}
\centerline{$^1$ Only two lines of NI were observed on Febr. 12 1993}
\end{table*}

\subsection{Hydrogen line profiles}

\begin{figure*}
\centerline{\hbox{\psfig{figure=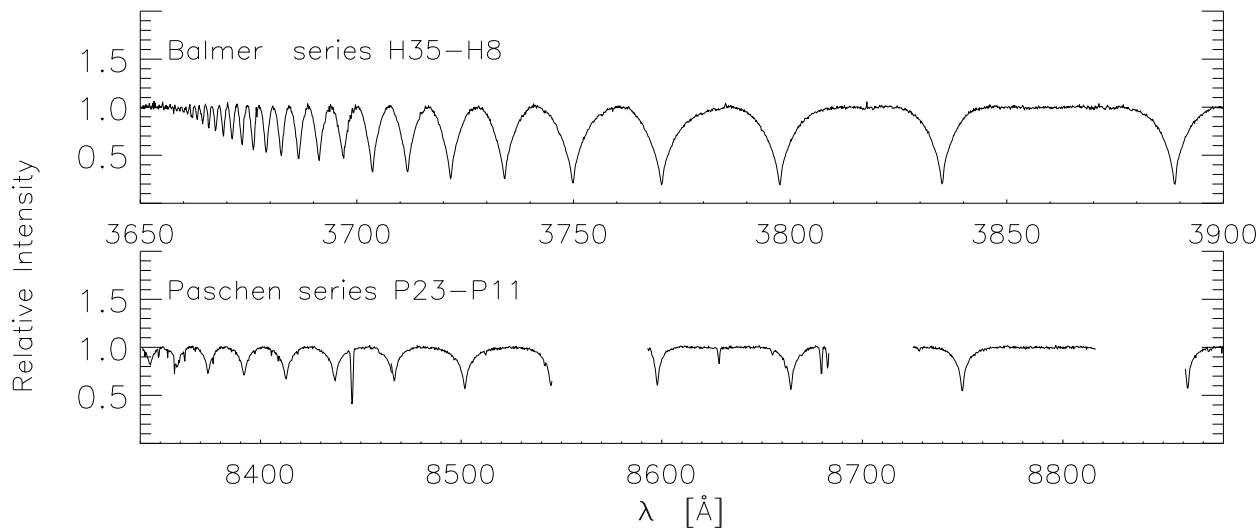,rheight=7cm,width=\textwidth}}}
\caption{The Balmer and Paschen series in the optical spectrum of HR~4049.
The complete optical spectrum is given in Fig.~9 available at CDS}
\label{art7fig-bapa}
\end{figure*}

The spectrum shows the Balmer lines from H$\alpha$
to H35 and the Paschen lines from P9 to P23. There are very
few stars known which can compete with HR~4049 in the number of Hydrogen
lines observed in the spectrum.  The part of the optical spectrum showing
the Balmer (H8-H35) and Paschen (P11-P23) series is shown in
Fig.~\ref{art7fig-bapa}. For a complete overview of the spectrum we
refer to Fig.~9 available at CDS.
Waelkens {\it et al. } (\cite{art7waelkenslamers}) were the first to note that
the shape of the H$\alpha$ profile changes with orbital and photometric
phase and attribute the H$\alpha$ variability to changes in the wind
structure caused by the presence of an unseen companion star.

\begin{figure*}
\centerline{\hbox{\psfig{figure=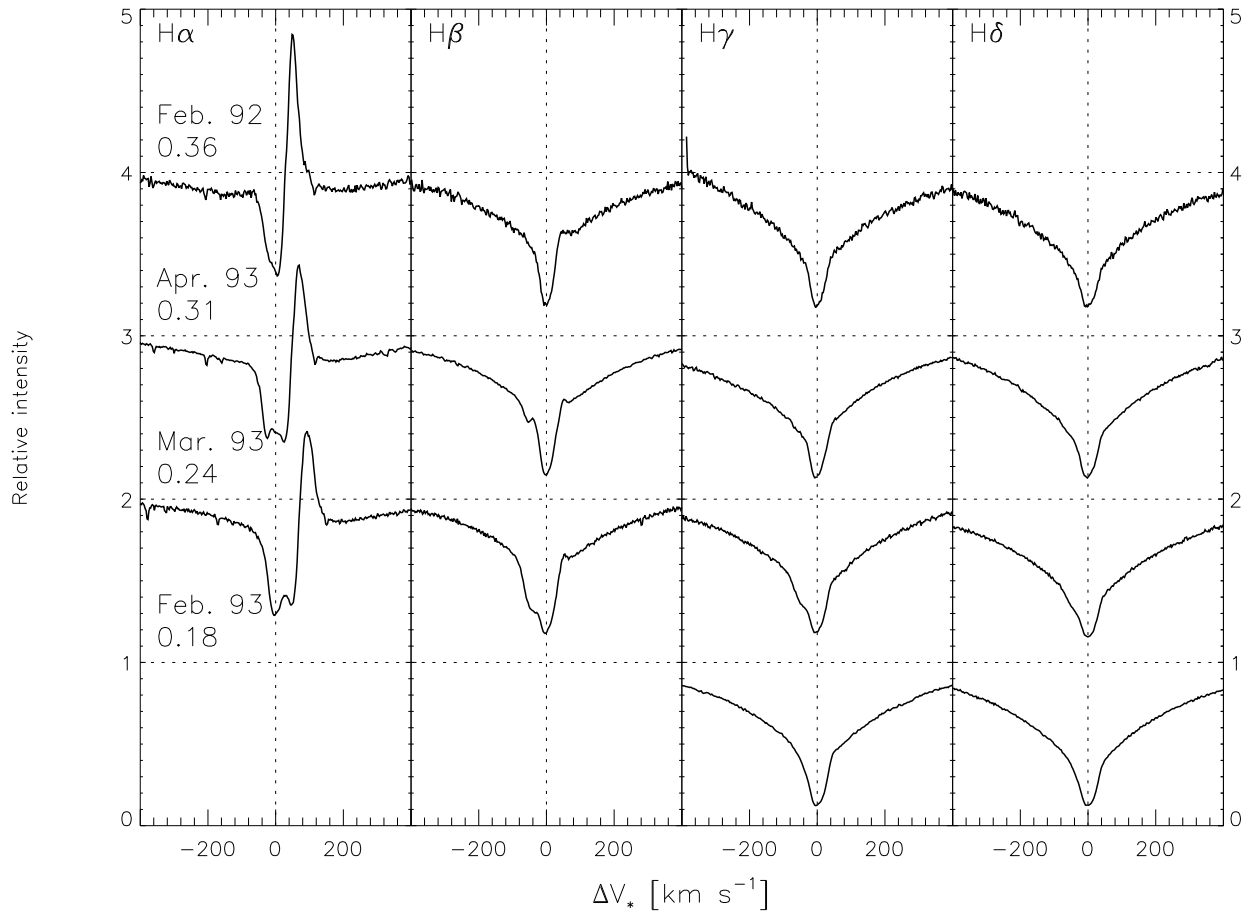,width=\textwidth}}}
\caption{The profiles of four Balmer lines at four different dates as
function of velocity with respect to the stellar velocity. The
spectra are plotted in order of orbital phase
($\phi=0.18$, 0.24, 0.31, 0.36). Lines which are not plotted
were not observed. Note that there is absorption on the red side of
the stellar velocity in the H$\alpha$ profile}
\label{art7fig-balmer}
\end{figure*}

In Fig.~\ref{art7fig-balmer} we have plotted the four lowest Balmer
lines on the dates we observed HR~4049.
We see that not only the line profile of
H$\alpha$, but also of H$\beta$ and H$\gamma$ show changes with time.
The first Balmer line which
has only a photospheric contribution is H$\epsilon$.
{}From the figure we see that the cores of the
Balmer lines are at the stellar velocity (corrected with
the velocity of CI, NI and OI). Surprisingly this is even
the case for the H$\alpha$ lines. From H$\alpha$ on April 1993 we find that
the absorption is from gas with velocities between -50 and +50~km~s$^{-1}$
with respect to the star, suggesting that both
infalling and outstreaming gas is observed at the same time.
Using the stellar parameters ($M_{\ast}$ and $R_{\ast}$
in Table~\ref{art7tab-hr4049}) we find
$v_{escape} =74$~km~s$^{-1}$. Slightly larger than the observed
outflow velocity of -50~km~s$^{-1}$.

With such a complex wind geometry it is not trivial to estimate the
mass-loss rate. We therefore tried to measure the Balmer progression
in order to see whether material is accreting or that we are looking at
wind material.

\begin{figure*}
\centerline{\hbox{\psfig{figure=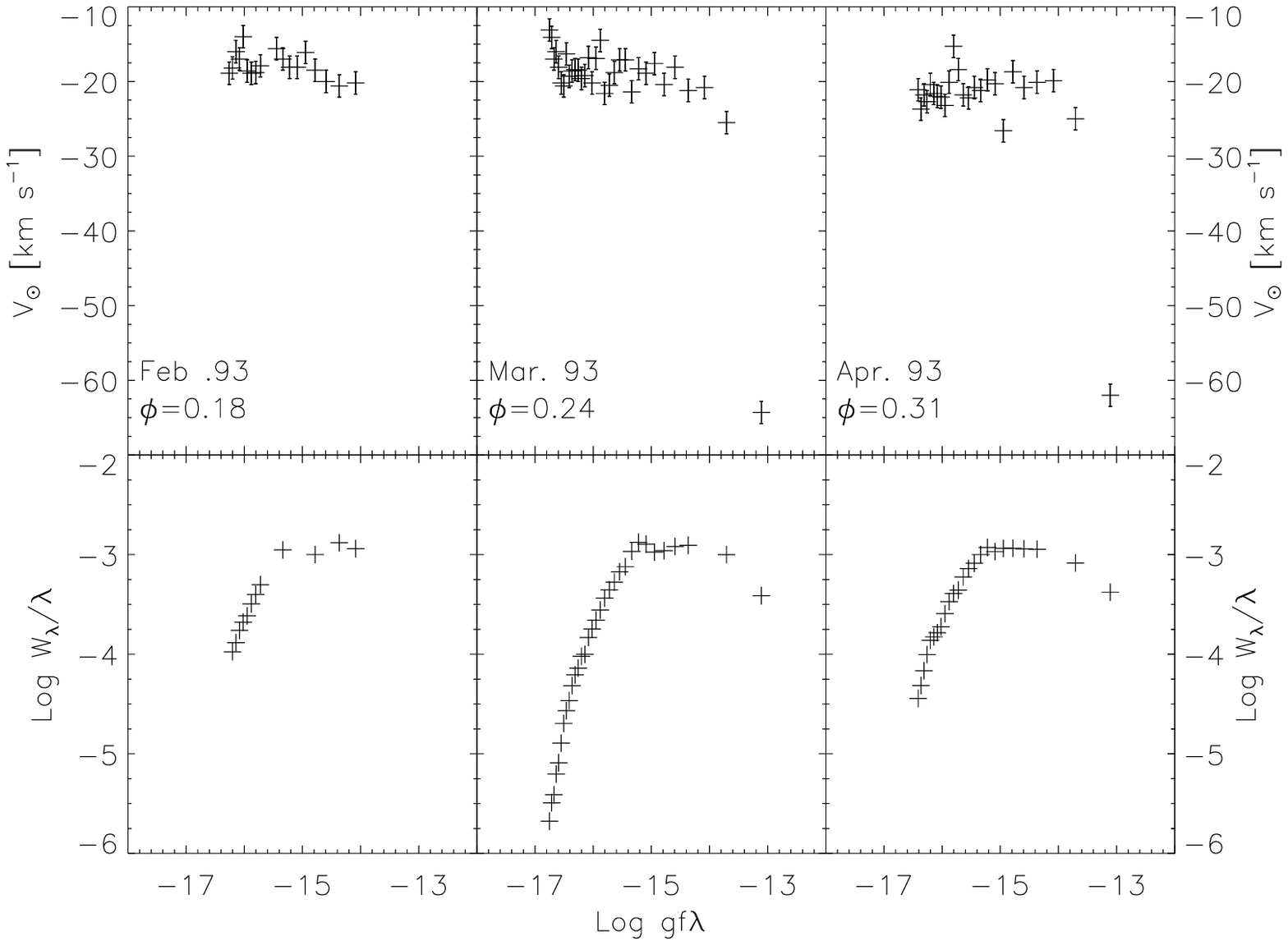,width=\textwidth}}}
\caption{The upper three plots show the Balmer progression at
three different phases during one orbital cycle.
The lower three plots show the curve of growth for the  Hydrogen Balmer lines}
\label{art7fig-balmercog}
\end{figure*}

The upper three plots of Fig.~\ref{art7fig-balmercog} do not clearly
show the presence of a Balmer progression. In February and March there seems
to be a small negative progression, in April there seems to be
a small positive progression. The H$\alpha$ and H$\beta$ lines
(two highest $\log gf \lambda$
points) are filled in by emission.
The lower three plots show
the curve of growth (CoG) on three different dates. The observations of
February 1992 were not included in Fig.~\ref{art7fig-balmercog}
as this spectrum
is from another orbital cycles and we know that the line profile variability
does not repeat itself exactly from one cycle to another
(Waelkens {\it et al. }
\cite{art7waelkenslamers}). The three CoG show the optically
thin part for $\log gf \lambda \leq -15.5$, the saturated part for
$-15.5 \leq \log gf \lambda \leq -14.0$. For $\log gf \lambda \geq -14.0$
(H$\alpha$ \& H$\beta$) the emission component
from the wind fills in the absorption and decreases the equivalent width.
For the higher Balmer lines (small $\log gf \lambda$) the lines blend
with each other and the continuum is no longer observed. This causes
an underestimate of the equivalent width.

\subsection{The resonance NaI D and CaII K lines}

\begin{figure*}
\centerline{\hbox{\psfig{figure=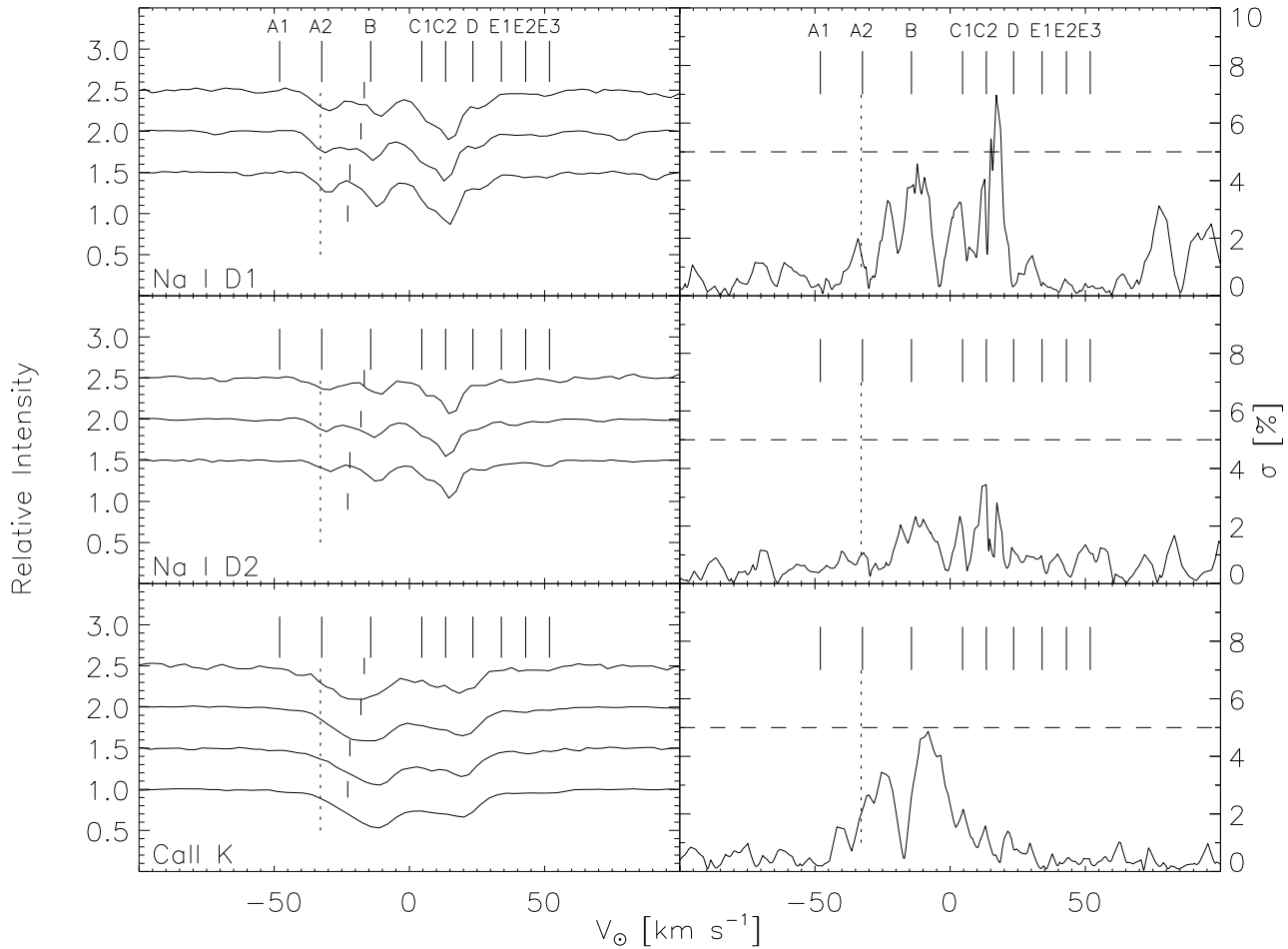,width=\textwidth}}}
\caption{Left: the resonance NaI D1 (top) \& D2 (middle) and CaII K (bottom)
lines are plotted on the velocity scale (from bottom to top:
Febr. 93, Mar. 93, Apr. 93 and Feb. 92).
The dashed line at -33~km~s$^{-1}$ is the system velocity
and the stellar velocity is marked by a small dash at the continuum level.
Right: the standard deviation of the four spectra as function of radial
velocity expressed in percentage of the continuum level}
\label{art7fig-naca}
\end{figure*}

\begin{figure*}
\centerline{\hbox{\psfig{figure=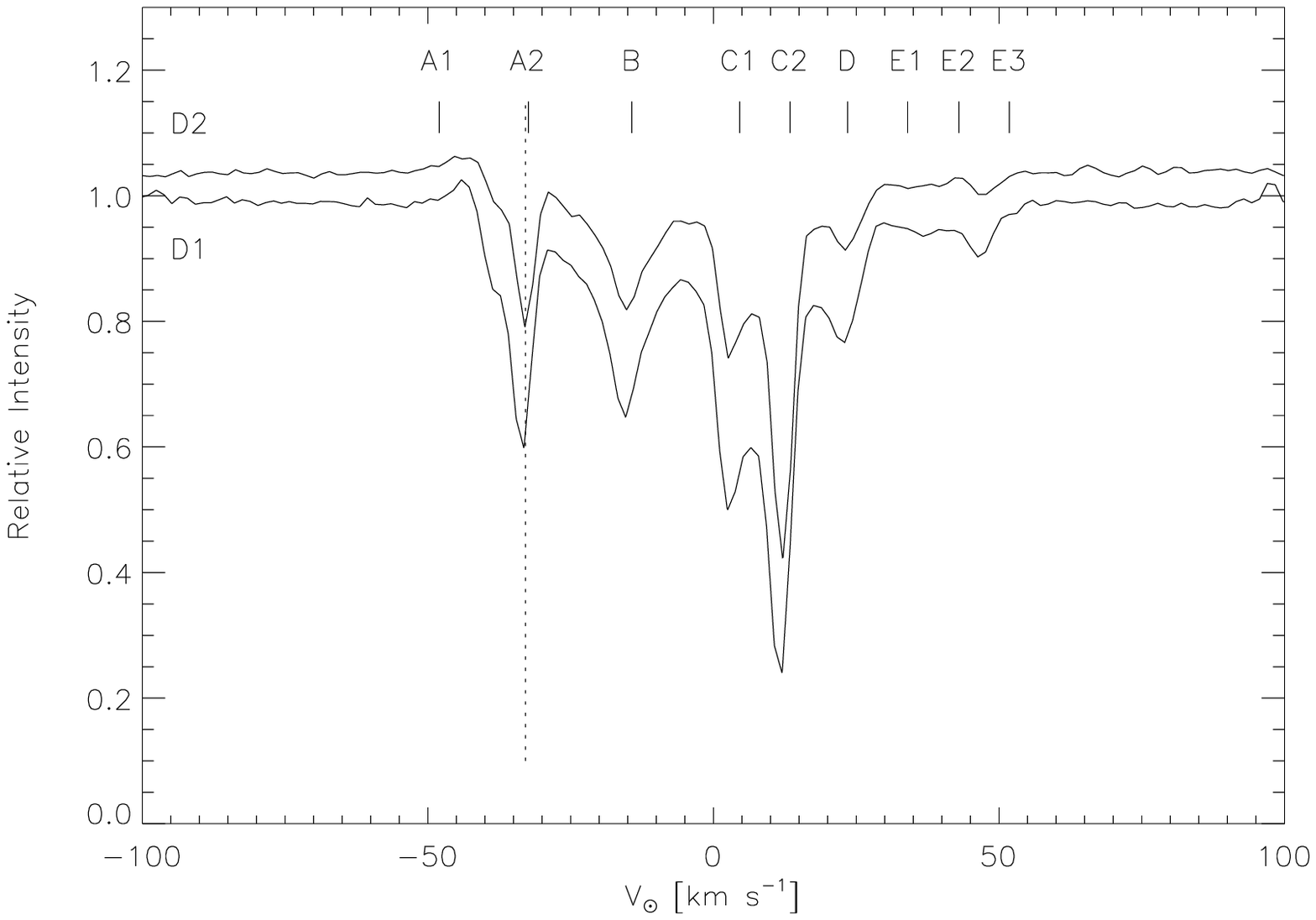,width=\textwidth}}}
\caption{NaI D1 \&\ D2 lines observed on Jan. 12 1986 with CAT/CES
when the star is closest to the circumsystem disk.
This plot clearly shows the presence of an emission component at the
blue edge of the NaI D1 \& D2 profiles (near A1).
The emission component is
not present in the UES/WHT spectra}
\label{art7fig-naijan86}
\end{figure*}

The strong resonance D lines of NaI and the H and K lines of CaII are the
best tracers in the optical spectrum of the diffuse interstellar medium
and circumstellar gas. From a study on the circumsystem
extinction of HR~4049 by Waelkens {\it et al. }
(\cite{art7waelkenslamers}) we know that the circumsystem
reddening changes with orbital phase.
We looked for a relation between the strength of the component and the
orbital phase.
Such a relation could be important in understanding the extreme
low metalicity of the photosphere of HR~4049 and could give
an answer whether gas is still accreting.
The left panel of
Fig.~\ref{art7fig-naca} shows the NaI D1 \& D2 lines and
the CaII K line obtained with WHT/UES and Fig.~\ref{art7fig-naijan86}
shows one of the CAT/CES spectra of the NaI D1 \& D2 lines.
These spectra show clearly the presence of five different
velocity components. To simplify
the discussion we have coded the main absorption
components in order of increasing
wavelength as A, B, C, D and E. Component A, C and E are blends of smaller
absorption lines.

To determine whether the NaI D1 \& D2 and CaII H \& K lines have
a photospheric, wind, circumstellar and an interstellar contribution
to the line profile, we will study the time variability of the
absorption components. In Table~\ref{art7tab-veliscs} the central
heliocentric velocities of all components have been listed in order
of velocities. Five strong (A...E) absorption components could be identified
of which three showed the presence of smaller absorption components. In
total we found nine components. Table~\ref{art7tab-veliscs} shows that none
of these absorption components shows radial velocity variation as
function of orbital phase. This means that there is no, or
a negligible photospheric contribution to the line profile and
the line forming region is stationary with respect to the system
velocity of HR~4049: inter- or circumstellar.

The first and the last spectrum are obtained
7.25~years apart which shows that no long term variations are observed.
Components C, D and E have positive velocities with respect to the
system velocity of -32.9~km~s$^{-1}$ suggesting that they are of interstellar
origin. Component A2 is on the system velocity and it is
therefore interesting to study the possibility that this line
is formed in the circumsystem disk. Component A1 is at negative velocities
with respect to the system velocity and could be due to the stellar
wind at a terminal velocity of  15~km~s$^{-1}$, much smaller than
the outflow velocity derived from the H$\alpha$ profile or the escape
velocity.
It is important to note that two of the spectra (CAS/CES)
which have been obtained
at the same orbital phase but from two successive cycles show an emission
component which fills in the A absorption component. In one
spectrum the emission component is even above the continuum level
at $v_{\odot} =-44$~km~s$^{-1}$. Such a strong change in line strength can only
occur if the line forming region is close to the star. If we look
at Fig.~\ref{art7fig-radvel} we note that the emission in the NaI D lines
occurs at apastron (radial velocity changes smoothly), thus
when the observed star is closest to the circumsystem disk. We therefore
interpret the emission as due to excitation of the circumsystem
disk by the radiation field of the star.

\begin{table*}
\caption{Velocities of absorption components in NaI D1 \& D2 and CaII H \& K
resonance lines}
\label{art7tab-veliscs}
\centerline{\rotate[l]{\begin{tabular}{|llllllllllll|}
\hline
$\phi$&Instr&line    & A1&A2 &B &C1 &C2&D &E1&E2&E3 \\
      &     &        & CS&CS &CS&IS &IS&IS&IS&IS&IS \\
\hline
      &     &        &   &   &   &  &  &  &  &  &   \\
0.10  &CAT  &D1$^{1}$&   &-34&-15& 3&12&23&  &  &   \\
0.15  &CAT  &D1$^{2}$&   &-33&-15& 4&12&23&37&47&   \\
0.24  &UES  &D1      &   &-29&-12& 6&15&27&  &  &51 \\
0.31  &UES  &D1      &   &-32&-14& 4&14&26&  &40&50 \\
0.90  &CAT  &D1      &-49&-34&-15& 3&11&22&  &46&52 \\
      &     &        &   &   &   &  &  &  &  &  &   \\
0.10  &CAT  &D2$^{1}$&   &-33&-15& 3&12&23&  &  &   \\
0.15  &CAT  &D2$^{1}$&   &-34&-15& 4&11&22&37&46&   \\
0.24  &UES  &D2      &   &-30&-12& 6&15&26&  &41&50 \\
0.31  &UES  &D2      &   &-31&-13& 5&14&25&  &  &50 \\
0.90  &CAT  &D2      &-54&-35&-16& 3&11&23&37&42&   \\
      &     &        &   &   &   &  &  &  &  &  &   \\
0.15  &CAT  &K       &   &   &-16& 4&15&  &37&48&   \\
0.18  &UES  &K       &-45&-30&-14& 5&15&22&32&43&53 \\
0.24  &UES  &K       &   &-33&-13& 8&19&  &30&43&56 \\
0.31  &UES  &K       &-42&-29&-16& 7&18&24&  &40&52 \\
0.88  &CAT  &K       &   &-39&-14& 3&15&  &  &  &   \\
      &     &        &   &   &   &  &  &  &  &  &   \\
0.24  &UES  &H$^{3}$ &   &   &-14&  &17&  &  &  &53 \\
      &     &        &   &   &   &  &  &  &  &  &   \\
\multicolumn{3}{|l}{$\overline{v_{\odot}}$ [km~s$^{-1}$]}
                 &$-48\pm3$
                     &$-32.5\pm0.7$
                         &$-14.3\pm0.3$
                              &$4.5\pm0.4$
                                &$14.1\pm0.6$
                                    &$23.8\pm0.8$
                                      &$35\pm1$
                                         &$43\pm1$
                                            &$51.9\pm0.7$\\
\multicolumn{3}{|l}{$\delta  v_{\rm system}$ [km~s$^{-1}$]}
                 &-15
                     &0.4
                         &18.6
                              &37.4
                                &47.0
                                    &56.7
                                      &68
                                         &76
                                            &85\\
      &     &        &   &   &   &  &  &  &  &  &   \\
\hline
\multicolumn{12}{l}{1: A1/A2 filled in by emission;
2: emission at -44~km~s$^{-1}$; 3: blended with H$\epsilon$} \\
\end{tabular}}}
\end{table*}

In Table~\ref{art7tab-ewiscs} the ratio of the observed equivalent width
for each component is given relative to the equivalent width as measured
on March 1993. The estimate of the error on
the equivalent width is less than 10 \%, but as different
instrument are used and the equivalent width are measured by different persons
using different techniques we will take a pessimistic estimate of
the error of 25~\%, meaning that all entries
in Table~\ref{art7tab-ewiscs} with values $\leq 0.7$ or $\geq 1.2$ indicate
that the equivalent width has significantly changed. Component E has
not been listed because the equivalent width is polluted by
a telluric H$_{2}$O lines, component C and D are treated as
one absorption component as they are severely blended with each other.

\begin{table}
\caption{Equivalent width of absorption components
in NaI D1 \& D2 and CaII H \& K
resonance lines relative to the March 1993 ($\phi =0.24$) observation}
\label{art7tab-ewiscs}
\centerline{\begin{tabular}{lllllll}
\hline
$\phi$&instr&line&A  &B  &C+D&remark      \\
      &     &    & CS&CS & IS&            \\
\hline
      &     &    &   &   &   &            \\
0.10  &CAT  &D1  &0.7&0.4&0.8&            \\
0.15  &CAT  &D1  &0.8&0.4&0.8&            \\
0.24  &UES  &D1  &1.0&1.0&1.0&$\equiv 1.0$\\
0.31  &UES  &D1  &1.0&0.7&1.0&            \\
0.90  &CAT  &D1  &3.0&0.3&0.9&            \\
      &     &    &   &   &   &            \\
0.10  &CAT  &D2  &0.8&0.5&0.9&            \\
0.15  &CAT  &D2  &0.8&0.5&0.9&            \\
0.24  &UES  &D2  &1.0&1.0&1.0&$\equiv 1.0$\\
0.31  &UES  &D2  &1.0&0.7&1.0&            \\
0.90  &CAT  &D2  &2.4&0.2&0.8&            \\
      &     &    &   &   &   &            \\
      &     &    &\multicolumn{4}{l}{A-D} \\
0.11  &CAT  &K   &0.8&   &   &            \\
0.15  &CAT  &K   &1.0&   &   &            \\
0.18  &UES  &K   &1.2&   &   &            \\
0.24  &UES  &K   &1.0&   &   &$\equiv 1.0$\\
0.31  &UES  &K   &1.1&   &   &            \\
0.88  &CAT  &K   &0.9&   &   &            \\
      &     &    &   &   &   &            \\
\hline
\end{tabular}}
\end{table}

This table shows that the C+D component shows no variations in the
equivalent width over the orbital phase, but the A and B component do
vary by a factor three in equivalent width! An extreme situation
occurs for NaI D1 at  $\phi =0.90$ where
the A component has increased in equivalent
width by a factor three while the equivalent width of
component B has decreased by a factor three.
There is also a remarkable difference between the CAT/CES and UES/WHT
spectra: in the first the component A is stronger than B,
where in the second B is stronger than A
(compare Fig.~\ref{art7fig-naca} and \ref{art7fig-naijan86}).
However this  data set is not suitable for a study on the exact dependence
of the equivalent width on orbital phase, but is shows
that the A and B component of NaI D1 \& D2 lines are variable and are
therefore not of interstellar origin.

Table~\ref{art7tab-ewiscs} also shows the ratio of equivalent width of the
CaII K lines at different dates with the equivalent width
as measured on March 1993.
No significant  variations are found.

We have seen that components A (A1 \& A2)
and B change with time (and probably orbital phase).
It is interesting to see whether the uniform data set of UES/WHT spectra
show short time variation ($0.18 \leq \phi \leq 0.31$) at less than 25~\%.
The right panel of Fig.~\ref{art7fig-naca} shows the standard
deviation ($\sigma$) on the spectra as function of velocity
expressed in \% to continuum level. The pollution by
telluric water vapor shows up in the Na I D1 profile as sharp variations of
less than 4~\%. The NaI D2 is less polluted and does not show significant
variations. The CaII K has smoothly changed for
$v_{\odot} \leq 0$~km~s$^{-1}$,
but at a very low level.

We conclude that the components at C,  D and E are probably of
{\it interstellar} origin and the variable components A and B
are of circumstellar origin.
Component A might be from wind material with an outflow velocity of
only 15.1~km~s$^{-1}$. So far we have assumed that the A2 and B components
are separate absorptions, but it is also possible that A2/B is
one strong absorption with a central emission component. Such a
situation can occur in a case of mass transfer in a binary system.

\subsection{[OI] emission line}

The presence of the [OI] line ($\chi = 0.0$~eV)
is of interest for the study of the
circumstellar environment of HR~4049 as it is generally thought
that forbidden lines are formed in low density gas.
We therefore have looked in detail at the spectra for the presence of
[OI] at 6300~\AA. Although very weak, we argue that we have detected
the [OI] at 6300~\AA~ in emission at three different dates.

\begin{figure}
\centerline{\hbox{\psfig{figure=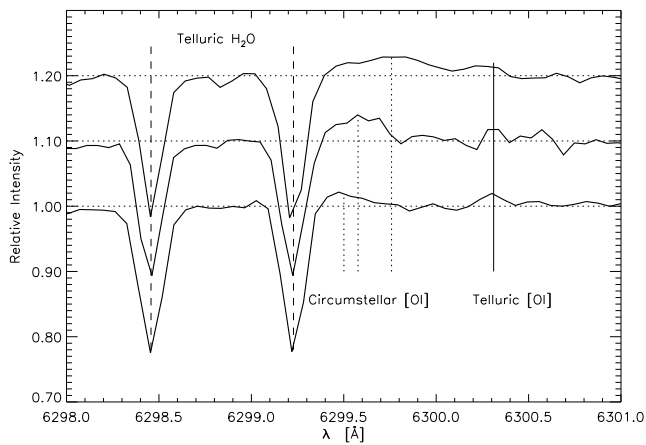,width=\columnwidth}}}
\caption{[OI]F in 3 different WHT/UES spectra of HR~4049
(from bottom to top: Feb93,  Mar93 and Apr93)}
\label{art7fig-oi}
\end{figure}

In Fig.~\ref{art7fig-oi} we have plotted the observed spectrum around
6300~\AA~ at three different phases of one cycle. We did not correct for
the earth motion, nor for the solar motion to show clearly the presence
of [OI] in emission. The two ``strong'' absorption lines are telluric
H$_{2}$O, while the emission lines at 6300.3~\AA~ is the
[OI] 6300.311~\AA~ telluric line. These lines do not vary in position.
A careful examination of this part
of the spectrum shows that the emission around 6299.6~\AA~ varies in
position. This excluded the line from being telluric.
Table~\ref{art7tab-oi}
give the heliocentric corrected velocity of the [OI] lines for
the three different dates.

\begin{table}
\caption{Radial velocities (heliocentric) of the observed [OI] emission
line at 6300~\AA~ at different orbital phases. The system velocity
is -32.9~km~s$^{-1}$}
\label{art7tab-oi}
\centerline{\begin{tabular}{lcc}
\hline
\multicolumn{1}{c}{Date} & $\phi$ & $v_{rad}$\\
 & & [km~s$^{-1}$] \\
\hline
            &      &          \\
Feb. 12 1993& 0.18 & $-25.6$  \\
Mar.  8 1993& 0.24 & $-31.0$  \\
Apr.  6 1993& 0.31 & $-32.5$  \\
            &      &          \\
Average     &      & $-30\pm2$\\
            &      &          \\
\hline
\end{tabular}}
\end{table}

The [OI]  emission lines does not follow the orbital motion of the
the CI, NI and OI lines, which excludes it from being photospheric.
If we compare
the [OI]  velocity of $-30\pm2$~km~s$^{-1}$ with the system velocity
of $-32.9\pm1.0$~km~s$^{-1}$ found from an analysis of the orbital
 motion of the star
(Van Winckel {\it et al. }
\cite{art7winckel}) we see that the [OI]  emission is
at the system velocity.
{}From the March 1993 spectrum (with the highest signal-to-noise-ratio)
we find a full-width-full-maximum of 20~km~s$^{-1}$, but the expansion
velocity derived from the
H$\alpha$ profile is of the order of 50~km~s$^{-1}$ whereas the
escape velocity is 75~km~s$^{-1}$. [OI]  should have a $FWFM$
of 100~km~s$^{-1}$
if it is formed in the stellar wind. As we only observe a $FWFM$
of 20~km~s$^{-1}$ the [OI]  it is not formed in the stellar wind.

Lambert {\it et al. } (\cite{art7lamberthinkle}) found the CO first overtone
in absorption at a velocity of $-33.1\pm0.2$~km~s$^{-1}$ with a
$T_{\rm excitation} =300 \pm 100$~K. This CO cannot be of photospheric
origin because the $T_{\rm exc}$ is too low, nor can it be formed
in the wind as it is observed at the system velocity.
Both the detection of [OI]  and CO first overtone can be explained
by the model of HR~4049 proposed by Waelkens {\it et al.}
(\cite{art7waelkens}) in
which a circumsystem disk revolves around the binary.
Assuming a Keplerian orbit
of the circumsystem disk around HR~4049
with $v \sin i=10$~km~s$^{-1}$
and taking the most likely $\sin i= \frac{1}{2} \sqrt{2}$
with the stellar parameters (Table~\ref{art7tab-hr4049})
we find a distance of the [OI] disk of only $r_{\rm [OI]} = 13 R_{\ast}$.

The distance from the circumsystem disk to HR~4049 can be estimated
in another way
by fitting the near-IR excess to an optically thin dust model.
This yields a dust inner radius of $r_{\rm inner} = 28 R_{\ast}$
(Lamers {\it et al. } \cite{art7lamerswaters}),  of the same order
as the radius found from the [OI] emission line. Taking the
dust inner radius and a Keplerian orbit yields a velocity
of 10~km~s$^{-1}$ of the disk material. To fit the [OI] velocity to
the dust inner radius the system has to be nearly edge-on
($i = 90^{o}$). This
explains the detection of CO first overtone in absorption  at the
system velocity (in the disk)
and the strong varying CS absorption
(Waelkens {\it et al. } \cite{art7waelkenslamers}).

One of the puzzling questions of HR~4049 and related objects  is the
mass and spectral type of the unseen companion star. Taking
$i = 90^{o}$ and the stellar parameters from Table~\ref{art7tab-hr4049}
we find a mass of the companion star of 0.56~M$_{\odot}$: a White Dwarf
or a low-mass main sequence star.

\section{Discussion}
\label{art7sec-dis}

\subsection{Mass-loss rate from asymmetric CI lines}

The asymmetry in the CI of Fig.~\ref{art7fig-clines} gives us a tool
to determine the mass-loss rate.  From the four profiles presented we
will not use $\lambda = 9078$~\AA~ because the continuum level
is not accurate enough. For the lines at $\lambda=5380$, 5052 and 4932~\AA~
we find that the extra absorption on the blue wing has an
equivalent width of respectively 9, 11 and 6~m\AA~ with
and estimated error of 3~m\AA. Taking the
$\log gf$ value from Table~11 and a statistical weight of
$g_{i}=3$ for all
three transitions we find column densities of respectively
$7.4 \times 10^{12}$, $6.5 \times 10^{12}$ and $5.8 \times 10^{12}$ cm$^{-2}$.
All three
transition are from the same lower level ($\chi=7.68$~eV) which allows
averaging of the column densities to $6.6\pm2.0 \times 10^{12}$ cm$^{-2}$.
We did not observed any CII lines
and neglected the CII contribution to the carbon abundance.
The wind temperature
is in the range $0.5 T_{\rm eff} \leq  T_{\rm wind} \leq  0.8 T_{\rm eff}$
with $T_{\rm eff}=7500$~K and we will adopt the upper limit
of $0.8 T_{\rm eff} = 6000$~K
meaning that we determine a minimum mass-loss rate.
Taking the
analytic approximation of the
partition function of CI from Gray (\cite{art7gray})
of $U(T_{\rm wind})=0.975$ while assuming LTE
we find a total column density for carbon of
$N_{C} =1.8 \pm0.3 \times 10^{19}$~cm$^{-2}$ and
$N_{H} =8.5 \pm0.2 \times 10^{22}$~cm$^{-2}$ taking the
carbon abundance of HR~4049 from Lambert {\it et al. }
(\cite{art7lamberthinkle}) of
[C/H]=-0.2.

In order to determine the mass-loss
\.{M}  we have to make an assumption about the
depth of the line forming region.  The terminal
velocity is not well known from H$\alpha$ because of the complex profile
and we use the relation of a late B-type supergiant of
$v_{\infty} =1.1 v_{\rm escape}$ found by Lamers {\it et al. }
(\cite{art7lamersetal}).
Taking the stellar parameters from Table~\ref{art7tab-hr4049}
we find an escape velocity of $v_{\rm escape} =75$~km~s$^{-1}$ and
a terminal velocity of the wind of $v_{\infty} =83$~km~s$^{-1}$.
In Sect.~\ref{art7subsec-cno} we have shown that the wind component
of the CI line profiles occurs at outflow velocities between
$v_{1} =10$~km~s$^{-1}$
and  $v_{2} = 40$~km~s$^{-1}$. With
the velocity law adopted (Eq.~\ref{art7eq-betalaw})  and a given $\beta$
(Table~\ref{art7tab-massloss}) this gives the geometrical distance of
the line forming region to the star. For $\beta =0.5$ we find
that the wind component is formed between $R_{1} =1.01 R_{\ast}$ and
$R_{2} =1.30 R_{\ast}$.

\begin{equation}
\label{art7eq-betalaw}
V(r) = V_{\infty}~ \left( 1 - \frac{R_{\ast}}{r} \right) ^{\beta}
{\rm ~[km~ s^{-1}]}
\end{equation}

\begin{table}
\caption{Mass-loss rates derived from the asymmetry of CI lines}
\label{art7tab-massloss}
\centerline{\begin{tabular}{llll}
\hline
       &          &          &                                \\
$\beta$& $r_{1}$  & $r_{2}$  &\.{M} ($T_{\rm wind}=6000$~K)      \\
       & [$R_{\ast}$]& [$R_{\ast}$]& [$M_{\odot}$ yr$^{-1}$]        \\
\hline
       &          &          &                                \\
0.5    & 1.01     & 1.30     & $9\pm3\times10^{-7}$           \\
1.0    & 1.13     & 1.93     & $4\pm3\times10^{-7}$           \\
2.0    & 1.53     & 3.27     & $4\pm3\times10^{-7}$           \\
       &          &          &                                \\
\hline
\end{tabular}}
\end{table}

The lower limit on the
mass-loss rates found (Table~\ref{art7tab-massloss}) is
several times $10^{-7}$~M$_{\odot}$~yr$^{-1}$ and
from this we adopt a post-AGB
mass-loss rate of HR~4049 of \.{M}$=6 \pm 4
\times 10^{-7}$~M$_{\odot}$~yr$^{-1}$.
This is the first time that the post-AGB
mass-loss rate has been determined  for HR~4049 and it shows that
HR~4049 has a significant stellar wind. Monitoring the
CI, NI and OI lines profiles over the orbital phase could give an answer
about the mass-loss rate as a function of orbital phase.

\section{Conclusion}
\label{art7sec-con}

With this paper we present a line identification list
of the optical spectrum of HR~4049 from 3650~\AA~ to 10850~\AA~. We
have identified 48 lines from HI and 116 lines from neutral CI, NI  and OI,
confirm the detection of HeI at 4471~\AA~ and present the first
detection of [OI] 6300~\AA~ emission at the system velocity.

Variability of the lines profiles is observed in H$\alpha$, H$\beta$ and
H$\gamma$,
but surprisingly also in the A \& B components of the resonance
NaI D1 \& D2 lines and the CaII K line.
At different orbital
phases the line-of-sight traces different circumstellar material
and results in a change of lines profile.
The emission in the NaI D lines is interpreted as due to excitation of
disk material by the radiation field of the star during apastron
(the star is closest to the CS disk).
Evidence for this model can be obtained
by simultaneously observing the NaI D lines, H$\alpha$ and the UV
(circumstellar reddening).

The discussion on the CI, NI and OI lines profiles shows that the stronger
C lines are asymmetric. The asymmetry can be explained as due
to a stellar wind with a mass-loss rate of
\.{M}$=6 \pm 4 \times10^{-7}$~M$_{\odot}$~yr$^{-1}$.
This is
the first time that the post-AGB mass-loss rate of HR~4049
has been determined.

The detection of [OI] in emission at the system velocity with
a Full-Width-Full-Maximum of only 20~km~s$^{-1}$ is interpreted as
due to an edge-on disk. For an edge-on system we find
from the orbital parameters that
the secondary has a mass of 0.56~M$_{\odot}$. This can be
a White Dwarf or a low-mass main sequence star.

\acknowledgements
The authors are very grateful to Rens Waters, Christoffel Waelkens,
Norman Trams and Ren\'{e} Oudmaijer for
the stimulating and constructive discussions on this work.
Ton Schoenmaker is acknowledged
for his valuable contribution in reducing the UES spectra.
EJB was supported by grant no. 782-371-040 by ASTRON, which receives funds
from the Netherlands Organization for the Advancement of Pure Research
(ZWO). This research has made use of the SIMBAD database, operated at CDS,
Strasbourg, France.

\typeout{textheight=56pc}

\begin{figure*}
\centerline{\hbox{\psfig{figure=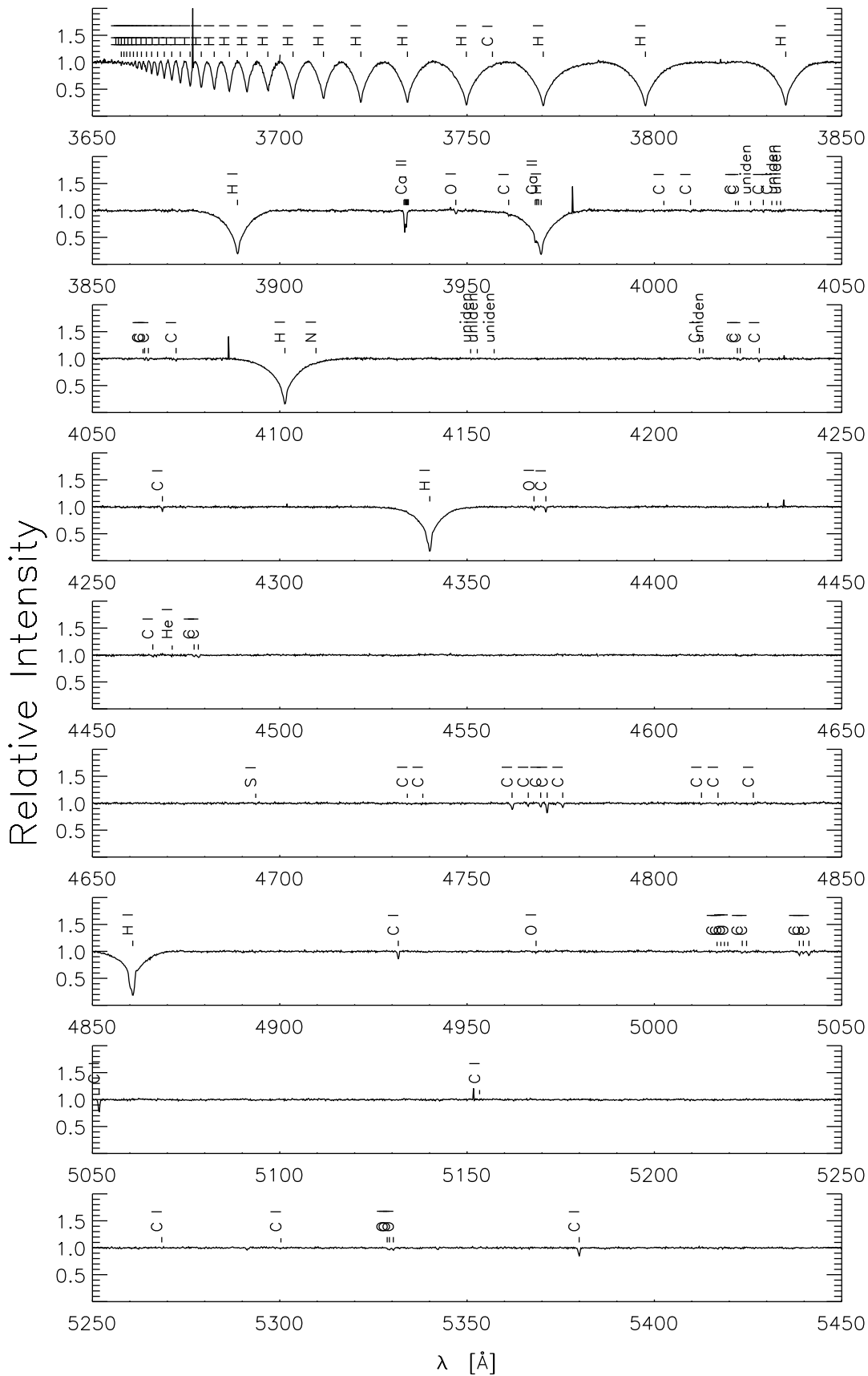,height=53pc}}}
\caption{The optical spectrum of HR~4049 from 3650~\AA~ to  10850~\AA}
\label{art7fig-spec}
\end{figure*}

\addtocounter{figure}{-1}
\begin{figure*}
\centerline{\hbox{\psfig{figure=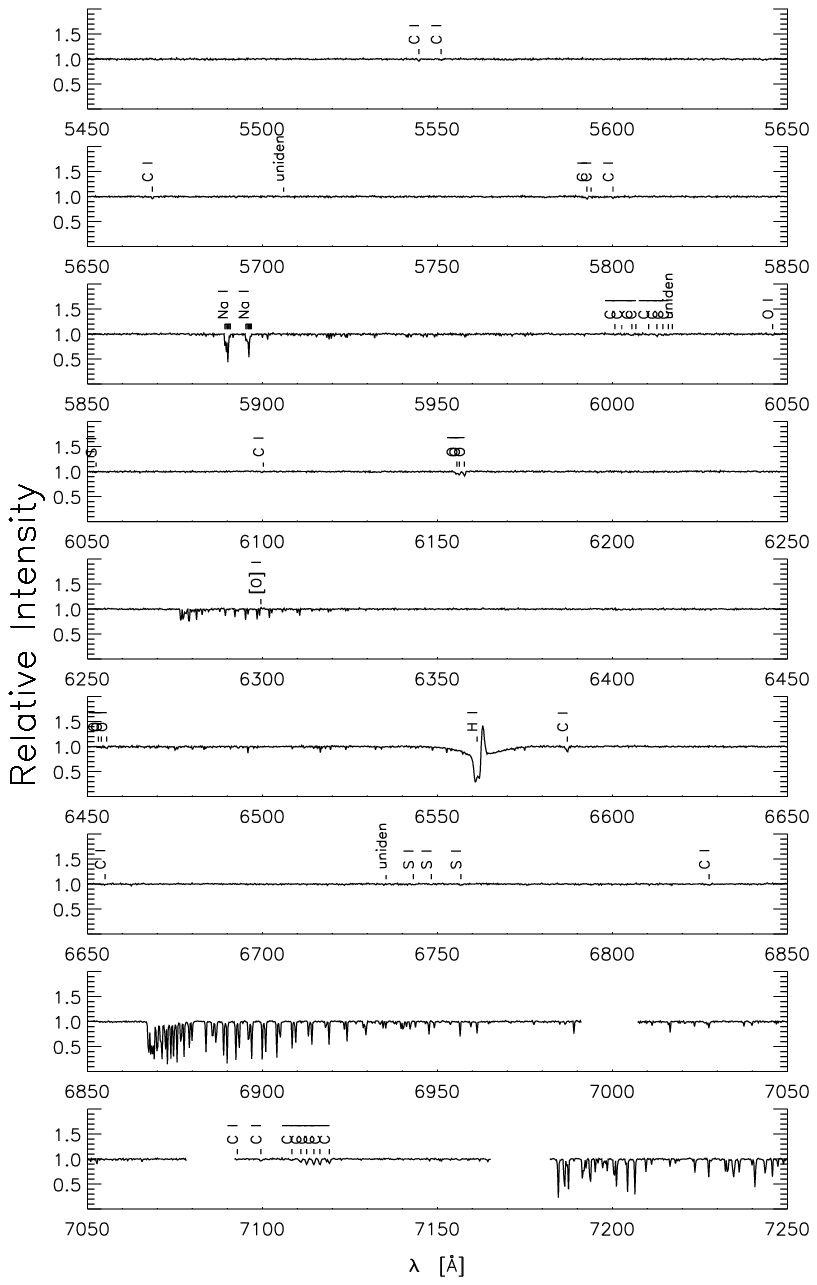,height=53pc}}}
\caption{continued}
\end{figure*}

\addtocounter{figure}{-1}
\begin{figure*}
\centerline{\hbox{\psfig{figure=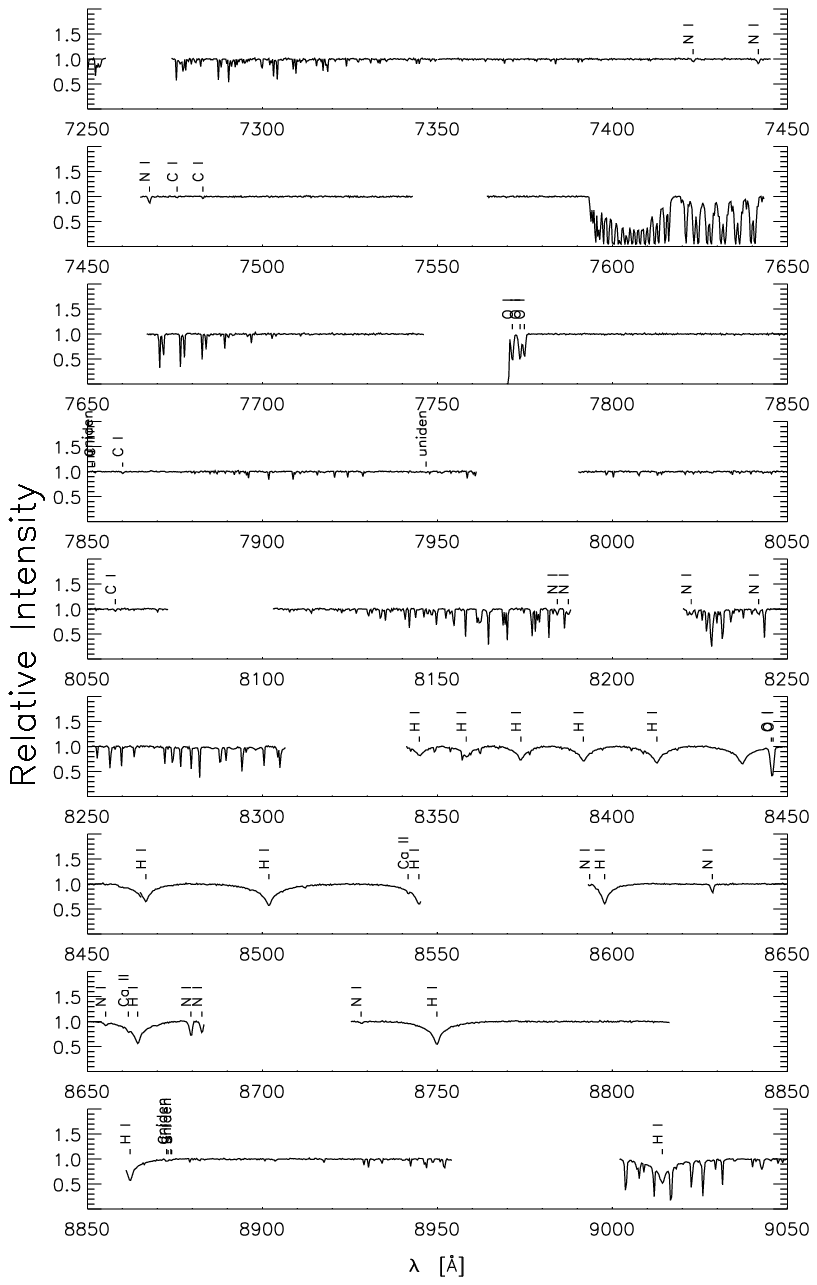,height=53pc}}}
\caption{continued}
\end{figure*}

\addtocounter{figure}{-1}
\begin{figure*}
\centerline{\hbox{\psfig{figure=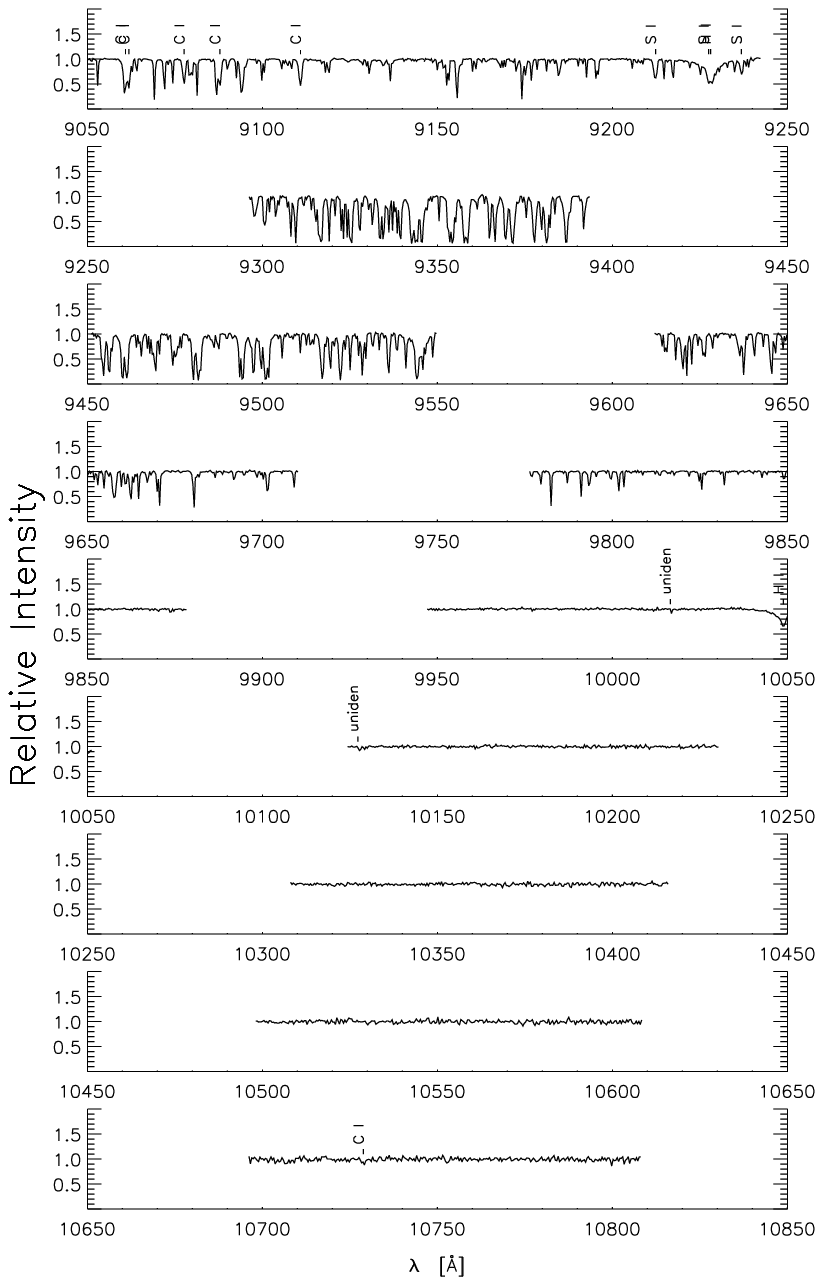,height=53pc}}}
\caption{continued}
\end{figure*}

\newcommand{\tabheaderartvii}{
                \addtocounter{table}{-1}
                \begin{centering}
                \begin{tiny}
                \begin{tabular}{|lcllrrrrrrrll|}
                \hline
                $\lambda_{\rm lab}$ &&
                multiplet           &
                $\chi$              &
                $\log gf$           &
                $\lambda_{\rm obs}$ &
                $D$                 &
                $W_{\lambda}$       &
                $FWHM$              &
                $v_{\rm rad}$       &
                run                 &
                $Q$                 &
                remarks             \\
                ~[\AA]               &&&
                ~[eV]                &&
                ~[\AA]               &
                ~[\%]                &
                ~[m\AA]              &
                ~[km/s]              &
                ~[km/s]              &&&
                                    \\
                \hline}

\newcommand{\tabendartvii}{
                     \hline
                     \end{tabular}
                     \end{tiny}
                     \end{centering}}

\begin{table*}
\caption{The line identification of HR~4049 from  3650~\AA~ to 10850~\AA}
\label{art7tab-id}
\tabheaderartvii
   3657.926  &     &   H I   7         &   10.15  & -3.189  &   3657.71  &  3.3
 &     8  &   21  &  -13  & 3.3  &  1  & H35
    \\
   3658.641  &     &   H I   7         &   10.15  & -3.151  &   3658.42  &  4.0
 &    12  &   23  &  -14  & 3.3  &  1  & H34
    \\
   3659.423  &     &   H I   6         &   10.15  & -3.112  &   3659.16  &  4.4
 &    14  &   25  &  -17  & 3.3  &  1  & H33
    \\
   3660.279  &     &   H I   6         &   10.15  & -3.072  &   3660.03  &  5.2
 &    23  &   32  &  -16  & 3.3  &  1  & H32
    \\
   3661.221  &     &   H I   6         &   10.15  & -3.030  &   3660.95  &  7.3
 &    30  &   34  &  -18  & 3.3  &  1  & H31
    \\
   3662.258  &     &   H I   6         &   10.15  & -2.987  &   3661.96  &  8.4
 &    47  &   41  &  -20  & 3.3  &  1  & H30
    \\
   3663.406  &     &   H I   6         &   10.15  & -2.942  &   3663.10  & 12.1
 &    74  &   38  &  -21  & 3.3  &  1  & H29
    \\
   3664.679  &     &   H I   6         &   10.15  & -2.896  &   3664.43  & 14.5
 &    99  &   44  &  -16  & 3.3  &  1  & H28
    \\
   3666.097  &     &   H I   5         &   10.15  & -2.848  &   3665.81  & 20.3
 &   125  &   43  &  -19  & 3.3  &  2  & H27
    \\
   3667.684  &     &   H I   5         &   10.15  & -2.799  &   3667.40  & 24.9
 &   177  &   51  &  -19  & 3.3  &  2  & H26
    \\
   3669.466  &     &   H I   5         &   10.15  & -2.747  &   3669.19  & 29.3
 &   227  &   54  &  -18  & 3.3  &  2  & H25
    \\
   3671.478  &     &   H I   5         &   10.15  & -2.693  &   3671.20  & 33.1
 &   266  &   73  &  -19  & 3.3  &  2  & H24
    \\
   3673.761  &     &   H I   5         &   10.15  & -2.637  &   3673.47  & 38.5
 &   350  &   69  &  -20  & 3.3  &  2  & H23
    \\
   3676.365  &     &   H I   4         &   10.15  & -2.578  &   3676.08  & 42.4
 &   366  &   78  &  -19  & 3.3  &  2  & H22
    \\
   3679.335  &     &   H I   4         &   10.15  & -2.517  &   3679.07  & 47.4
 &   542  &   83  &  -17  & 3.3  &  3  & H21
    \\
   3682.810  &     &   H I   4         &   10.15  & -2.452  &   3682.51  & 50.6
 &   660  &  101  &  -20  & 3.3  &  3  & H20, a
    \\
   3686.833  &     &   H I   4         &   10.15  & -2.384  &   3686.57  & 53.5
 &   811  &  107  &  -17  & 3.3  &  3  & H19, a
    \\
   3691.557  &     &   H I   4         &   10.15  & -2.312  &   3691.32  & 58.4
 &  1025  &  125  &  -14  & 3.3  &  3  & H18, a
    \\
   3697.154  &     &   H I   3         &   10.15  & -2.235  &   3696.83  & 62.2
 &  1350  &  138  &  -22  & 3.3  &  2  & H17, a
    \\
   3703.855  &     &   H I   3         &   10.15  & -2.154  &   3703.55  & 66.6
 &  1639  &  167  &  -20  & 3.3  &  3  & H16, a
    \\
   3711.973  &     &   H I   3         &   10.15  & -2.067  &   3711.69  & 67.4
 &  1964  &  200  &  -19  & 3.3  &  3  & H15, a
    \\
   3721.940  &     &   H I   3         &   10.15  & -1.974  &   3721.67  & 73.9
 &  2499  &  236  &  -17  & 3.3  &  2  & H14, a
    \\
   3734.370  &     &   H I   2         &   10.15  & -1.873  &   3734.10  & 74.1
 &  2824  &  254  &  -17  & 3.3  &  3  & H13, a
    \\
   3750.154  &     &   H I   2         &   10.15  & -1.764  &   3749.83  & 78.4
 &  4034  &  334  &  -21  & 3.3  &  3  & H12, a
    \\
   3757.042  &     &   C I   7.04      &    7.49  & -3.550  &   3756.80  &  4.5
 &    19  &   32  &  -15  & 3.3  &  1  & b
    \\
   3770.632  &     &   H I   2         &   10.15  & -1.644  &   3770.35  & 80.5
 &  5020  &  361  &  -18  & 3.3  &  1  & H11, a, wrong
    \\
   3797.900  &     &   H I   2         &   10.15  & -1.511  &   3797.61  & 80.7
 &  4843  &  347  &  -19  & 3.3  &  3  & H10, a
    \\
   3835.368  &     &   H I   2         &   10.15  & -1.362  &   3835.09  & 79.4
 &  4060  &  300  &  -18  & 3.3  &  3  & H9, a
    \\
   3889.051  &     &   H I   2         &   10.15  & -1.192  &   3888.73  & 79.7
 &  4265  &  297  &  -20  & 3.3  &  3  & H8, a, wrong
    \\
   3933.664  &     &  Ca II  1         &    0.00  &  0.150  &   3933.18  & 10.0
 &    28  &   20  &  -33  & 3.3  &  1  & A2, CS
    \\
   3933.664  &     &  Ca II  1         &    0.00  &  0.150  &   3933.43  & 43.8
 &   131  &   21  &  -13  & 3.3  &  2  & B, CS, b
    \\
   3933.664  &     &  Ca II  1         &    0.00  &  0.150  &   3933.71  & 25.0
 &    44  &   13  &    8  & 3.3  &  1  & C1, IS, b
    \\
   3933.664  &     &  Ca II  1         &    0.00  &  0.150  &   3933.86  & 32.1
 &    46  &   10  &   19  & 3.3  &  1  & C2, IS, b
    \\
   3933.664  &     &  Ca II  1         &    0.00  &  0.150  &   3933.99  &  6.0
 &    11  &   13  &   30  & 3.3  &  1  & E1, IS, b
    \\
   3933.664  &     &  Ca II  1         &    0.00  &  0.150  &   3934.18  &  6.8
 &    10  &   11  &   43  & 3.3  &  1  & E2, IS, b
    \\
   3933.664  &     &  Ca II  1         &    0.00  &  0.150  &   3934.34  &  3.3
 &     6  &   13  &   56  & 3.3  &  1  & E3, IS, b
    \\
   3947.295  &     &   O I   3         &    9.15  & -2.280  &   3947.01  &  5.3
 &    36  &   49  &  -17  & 3.3  &  1  & b
    \\
   3961.403  &     &   C I   17.03     &    7.68  & -2.140  &   3961.13  &  4.2
 &    13  &   23  &  -17  & 3.3  &  1  & b
    \\
   3968.470  &     &  Ca II  1         &    0.00  & -0.150  &   3968.22  & 13.8
 &    49  &   25  &  -14  & 3.3  &  1  & B, CS, b
    \\
   3968.470  &     &  Ca II  1         &    0.00  & -0.150  &   3968.64  &  6.7
 &    15  &   16  &   17  & 3.3  &  1  & C2, IS, b
    \\
   3968.470  &     &  Ca II  1         &    0.00  & -0.150  &   3969.11  &  3.4
 &    11  &   23  &   53  & 3.3  &  1  & E3, IS, b
    \\
   3970.074  &     &   H I   2         &   10.15  & -0.993  &   3969.78  & 81.6
 &  4800  &  305  &  -18  & 3.3  &  2  & H$\epsilon$
    \\
   4002.978  &  *  &   C I   17.02     &    7.68  & -3.060  &   4002.55  &  1.6
 &     4  &   17  &  -17  & 2.3  &  1  &
    \\
   4009.930  &     &   C I   17.01     &    7.68  & -2.780  &   4009.65  &  3.9
 &    18  &   29  &  -17  & 3.3  &  1  &
    \\
   4022.114  &  *  &   C I   7.01      &    7.48  & -2.770  &   4021.67  &  2.0
 &     8  &   28  &  -19  & 2.3  &  1  &
    \\
   4022.814  &  *  &   C I   7.01      &    7.48  & -2.650  &   4022.44  &  2.3
 &     9  &   28  &  -14  & 2.3  &  1  &
    \\
             &  *  &  unident          &          &         &   4025.68  &  2.2
 &    14  &   46  &       & 2.3  &  1  &
    \\
   4029.413  &     &   C I   7.01      &    7.49  & -2.190  &   4029.09  &  3.2
 &    15  &   29  &  -20  & 3.3  &  1  &
    \\
   4031.800  &  *  &   C I   7.01      &    7.49  & -2.790  &   4031.36  &  1.2
 &     5  &   28  &  -18  & 2.3  &  1  &
    \\
             &  *  &  unident          &          &         &   4032.67  &  1.9
 &     6  &   23  &       & 4.3  &  1  &
    \\
             &  *  &  unident          &          &         &   4033.71  &  1.1
 &     6  &   39  &       & 2.3  &  1  &
    \\
   4063.577  &  *  &   C I   7         &    7.48  & -3.782  &   4063.50  &  2.0
 &    11  &   39  &  -13  & 4.3  &  1  &
    \\
   4064.268  &     &   C I   7         &    7.48  & -3.430  &   4063.94  &  2.5
 &     7  &   17  &  -20  & 3.3  &  1  & b
    \\
   4065.243  &     &   C I   7         &    7.49  & -3.159  &   4064.95  &  3.8
 &    14  &   25  &  -17  & 3.3  &  1  &
    \\
   4072.643  &     &   C I   18.17     &          &         &   4072.35  &  3.0
 &    15  &   33  &  -17  & 3.3  &  1  &
    \\
   4101.737  &     &   H I   1         &   10.15  & -0.753  &   4101.39  & 84.3
 &  5082  &  272  &  -21  & 3.3  &  3  & H$\delta$
    \\
   4109.949  &     &   N I   10        &   10.69  & -1.200  &   4109.70  &  2.0
 &     9  &   31  &  -13  & 3.3  &  1  & b
    \\
             &  *  &  unident          &          &         &   4150.98  &  1.7
 &     7  &   27  &       & 2.3  &  1  &
    \\
             &  *  &  unident          &          &         &   4152.73  &  1.7
 &     7  &   27  &       & 4.3  &  1  &
    \\
             &  *  &  unident          &          &         &   4157.23  &  1.5
 &     6  &   27  &       & 2.3  &  1  &
    \\
   4212.342  &     &   C I   18.12     &          &         &   4212.06  &  4.6
 &    11  &   22  &  -16  & 3.3  &  1  &
    \\
             &  *  &  unident          &          &         &   4213.02  &  1.8
 &     5  &   18  &       & 4.3  &  1  &
    \\
   4222.466  &     &   C I   18.11     &          &         &   4222.16  &  2.0
 &     6  &   22  &  -17  & 3.3  &  1  &
    \\
   4223.360  &     &   C I   18.11     &          &         &   4222.93  &  3.3
 &    16  &   32  &  -26  & 3.3  &  1  & b
    \\
   4228.326  &     &   C I   17        &    7.68  & -2.794  &   4228.01  &  6.0
 &    25  &   28  &  -18  & 3.3  &  1  &
    \\
   4269.020  &     &   C I   16        &    7.68  & -2.542  &   4268.68  &  8.9
 &    31  &   23  &  -19  & 3.3  &  1  &
    \\
   4340.427  &     &   H I   1         &   10.15  & -0.447  &   4340.06  & 81.5
 &        &       &  -21  & 3.3  &  1  & H$\gamma$, half profile
    \\
   4368.242  &     &   O I   5         &    9.52  & -2.030  &   4367.88  &  6.5
 &    28  &   29  &  -21  & 3.3  &  1  & b
    \\
   4371.367  &     &   C I   14        &    7.68  & -2.333  &   4371.05  &  9.5
 &    38  &   26  &  -17  & 3.3  &  2  &
    \\
   4466.476  &     &   C I   18.07     &          &         &   4466.14  &  2.9
 &    15  &   32  &  -18  & 3.2  &  1  &
    \\
   4471.477  &  *  &  He I   14        &   20.87  &  0.050  &   4471.29  &  2.5
 &    21  &   52  &  -20  & 4.2  &  3  &
    \\
   4477.472  &     &   C I   18.06     &          &         &   4477.15  &  4.3
 &    15  &   22  &  -17  & 3.3  &  1  &
    \\
   4478.588  &     &   C I   18.06     &          &         &   4478.27  &  5.1
 &    32  &   42  &  -17  & 3.2  &  1  & b
    \\
   4694.113  &  *  &   S I   2         &    6.52  & -1.770  &   4693.60  &  2.4
 &     8  &   19  &  -18  & 2.2  &  1  &
    \\
   4734.260  &  *  &   C I   18.05     &    7.95  & -2.090  &   4734.08  &  2.6
 &    12  &   26  &  -19  & 4.2  &  1  &
    \\
   4738.461  &  *  &   C I   18.05     &    7.95  & -2.360  &   4738.20  &  1.7
 &    10  &   33  &  -24  & 4.2  &  1  & b
    \\
   4762.313  &     &   C I   6         &    7.48  & -2.530  &   4762.03  & 13.9
 &    87  &   35  &  -14  & 3.2  &  2  & b
    \\
   4766.672  &     &   C I   6         &    7.48  & -2.309  &   4766.34  &  5.7
 &    21  &   22  &  -17  & 3.2  &  1  &
    \\
   4770.027  &     &   C I   6         &    7.48  & -2.722  &   4769.66  &  7.1
 &    32  &   27  &  -18  & 3.2  &  2  &
    \\
%
   4771.742  &     &   C I   6         &    7.49  & -2.120  &   4771.38  & 19.4
 &    87  &   26  &  -18  & 3.2  &  3  &
    \\
   4775.897  &     &   C I   6         &    7.49  & -2.163  &   4775.56  &  9.4
 &    38  &   24  &  -17  & 3.2  &  2  &
    \\
   4812.920  &     &   C I   5         &    7.48  & -3.505  &   4812.58  &  1.7
 &    11  &   37  &  -17  & 3.2  &  1  &
    \\
   4817.373  &     &   C I   5         &    7.48  & -3.203  &   4817.01  &  3.5
 &    13  &   22  &  -18  & 3.2  &  1  &
    \\
   4826.796  &     &   C I   5         &    7.49  & -2.980  &   4826.41  &  1.7
 &     6  &   22  &  -19  & 3.2  &  1  &
    \\
   4861.279  &     &   H I   2         &   10.15  & -0.020  &   4860.79  & 82.1
 &  4861  &  127  &  -26  & 3.2  &  3  & H$\beta$, core
    \\
   4932.049  &     &   C I   13        &    7.68  & -1.844  &   4931.66  & 14.9
 &    62  &   24  &  -19  & 3.2  &  2  &
    \\
   4968.793  &     &   O I   14        &   10.69  &         &   4968.42  &  1.8
 &     7  &   21  &  -18  & 3.2  &  1  &
    \\
   5017.090  &  *  &   C I   18.03     &    7.95  & -0.500  &   5016.75  &  1.3
 &     4  &   28  &   -6  & 2.2  &  1  &
    \\
   5018.068  &  *  &   C I   18.03     &    7.95  & -2.000  &   5017.75  &  1.3
 &     7  &   51  &   -5  & 2.2  &  1  &
    \\
   5019.291  &  *  &   O I   13        &   10.74  & -1.900  &   5018.71  &  1.2
 &     3  &   16  &  -20  & 2.2  &  1  & b
    \\
   5020.218  &  *  &   O I   13        &   10.74  & -1.760  &   5019.62  &  0.9
 &     4  &   24  &  -21  & 2.2  &  1  & b
    \\
   5023.849  &     &   C I   18.02     &    7.95  & -2.400  &   5023.46  &  2.8
 &    16  &   31  &  -19  & 3.2  &  1  &
    \\
   5024.916  &     &   C I   18.02     &    7.95  & -2.700  &   5024.63  &  2.0
 &     7  &   41  &  -13  & 3.2  &  1  &
    \\
   5039.057  &     &   C I   18.01/4   &    7.95  & -2.000  &   5038.70  &  6.8
 &    41  &   34  &  -17  & 3.2  &  1  & b
    \\
   5040.134  &     &   C I   18.01     &    7.95  & -2.500  &   5039.77  &  2.8
 &    17  &   29  &  -17  & 3.2  &  1  &
    \\
   5041.660  &     &   C I   18.01/4   &    7.95  & -2.500  &   5041.28  &  7.8
 &    47  &   34  &  -18  & 3.2  &  1  & b
    \\
   5052.167  &     &   C I   12        &    7.68  & -1.648  &   5051.78  & 24.2
 &   115  &   26  &  -19  & 3.2  &  3  &
    \\
   5153.567  &  *  &   C I   26.22     &    8.65  & -4.570  &   5153.37  &  1.9
 &     9  &   26  &  -19  & 4.2  &  1  &
    \\
   5268.948  &     &   C I   22.1      &    8.54  & -2.280  &   5268.52  &  2.1
 &    10  &   26  &  -20  & 3.2  &  1  &
    \\
   5329.099  &     &   O I   12        &   10.74  & -1.730  &   5328.70  &  2.6
 &    12  &   24  &  -18  & 3.2  &  1  &
    \\
   5329.690  &     &   O I   12        &   10.74  & -1.410  &   5329.29  &  4.1
 &    16  &   21  &  -18  & 3.2  &  1  & b
    \\
   5300.550  &  *  &   C I   26.19     &    8.65  & -4.190  &   5300.31  &  2.1
 &    12  &   30  &  -22  & 4.2  &  1  & b
    \\
   5330.741  &     &   O I   12        &   10.74  & -1.120  &   5330.34  &  4.5
 &    24  &   28  &  -18  & 3.2  &  1  &
    \\
   5380.337  &     &   C I   11        &    7.68  & -1.842  &   5379.94  & 14.3
 &    77  &   28  &  -18  & 3.1  &  2  &
    \\
   5545.055  &     &   C I   26.14/15  &    8.64  & -2.510  &   5544.69  &  4.3
 &    16  &   26  &  -15  & 3.1  &  1  & b
    \\
\tabendartvii
\end{table*}

\begin{table*}
\caption{continued}
\tabheaderartvii
   5551.576  &     &   C I   26.14     &    8.64  & -2.030  &   5551.04  &  1.6
 &    24  &   27  &  -25  & 3.2  &  1  &
    \\
   5668.943  &     &   C I   22.06     &    8.54  & -2.430  &   5668.54  &  4.7
 &    21  &   24  &  -17  & 3.1  &  1  &
    \\
             &  *  &  unident          &          &         &   5706.02  &  1.7
 &     7  &   19  &       & 4.2  &  1  &
    \\
   5793.120  &     &   C I   18        &    7.95  & -2.157  &   5792.70  &  5.0
 &    33  &   33  &  -18  & 3.1  &  1  &
    \\
   5794.473  &  *  &   C I   18        &    7.95  & -2.900  &   5793.91  &  1.0
 &     9  &   43  &  -15  & 2.2  &  1  &
    \\
   5800.594  &     &   C I   18        &    7.95  & -2.440  &   5800.15  &  2.6
 &    19  &   23  &  -19  & 3.1  &  1  &
    \\
   5889.953  &     &  Na I   1         &    0.00  &  0.110  &   5889.28  & 24.6
 &    49  &   10  &  -30  & 3.1  &  3  & A2, CS
    \\
   5889.953  &     &  Na I   1         &    0.00  &  0.110  &   5889.62  & 42.9
 &    98  &   11  &  -12  & 3.1  &  3  & B,  CS
    \\
   5889.953  &     &  Na I   1         &    0.00  &  0.110  &   5889.99  & 38.9
 &    69  &    8  &    6  & 3.1  &  1  & C1, IS
    \\
   5889.953  &     &  Na I   1         &    0.00  &  0.110  &   5890.16  & 62.8
 &   111  &    8  &   15  & 3.1  &  1  & C2, IS
    \\
   5889.953  &     &  Na I   1         &    0.00  &  0.110  &   5890.38  & 20.8
 &    39  &    9  &   26  & 3.1  &  1  & D,  IS
    \\
   5889.953  &     &  Na I   1         &    0.00  &  0.110  &   5890.68  &  6.0
 &    10  &    8  &   41  & 3.1  &  1  & E2, IS
    \\
   5889.953  &     &  Na I   1         &    0.00  &  0.110  &   5890.85  &  6.8
 &    12  &    8  &   50  & 3.1  &  1  & D3, IS
    \\
   5895.923  &     &  Na I   1         &    0.00  & -0.190  &   5895.26  & 13.7
 &    27  &    9  &  -29  & 3.1  &  2  & A2, CS
    \\
   5895.923  &     &  Na I   1         &    0.00  & -0.190  &   5895.61  & 27.5
 &    58  &   10  &  -12  & 3.1  &  2  & B,  CS
    \\
   5895.923  &     &  Na I   1         &    0.00  & -0.190  &   5895.96  & 21.5
 &    38  &    8  &    6  & 3.1  &  1  & C1, IS
    \\
   5895.923  &     &  Na I   1         &    0.00  & -0.190  &   5896.14  & 48.0
 &    78  &    8  &   15  & 3.1  &  1  & C2, IS
    \\
   5895.923  &     &  Na I   1         &    0.00  & -0.190  &   5896.37  & 12.8
 &    23  &    8  &   27  & 3.1  &  1  & D,  IS
    \\
   5895.923  &     &  Na I   1         &    0.00  & -0.190  &   5896.57  &  5.0
 &     9  &    8  &   37  & 3.1  &  1  & E1, IS
    \\
   5895.923  &     &  Na I   1         &    0.00  & -0.190  &   5896.83  &  7.8
 &    14  &    8  &   51  & 3.1  &  1  & E2, IS
    \\
   6001.121  &     &   C I   26.07     &    8.64  & -2.070  &   6000.67  &  3.1
 &     5  &   26  &  -18  & 3.1  &  1  &
    \\
   6002.983  &  *  &   C I   26.08     &    8.65  & -2.050  &   6002.65  &  1.9
 &    10  &   25  &  -24  & 4.1  &  1  &
    \\
   6006.021  &     &   C I   26.08     &    8.65  & -3.290  &   6005.56  &  3.1
 &    16  &   24  &  -18  & 3.2  &  1  &
    \\
   6007.175  &     &   C I   26.07     &    8.64  & -2.180  &   6006.70  &  1.8
 &     8  &   21  &  -19  & 3.2  &  1  &
    \\
   6010.675  &     &   C I   26.07     &    8.64  & -2.020  &   6010.30  &  2.0
 &    10  &   23  &  -14  & 3.1  &  1  &
    \\
   6013.166  &     &   C I   26.07/06  &    8.65  & -1.370  &   6012.70  &  5.6
 &    39  &   32  &  -19  & 3.1  &  1  & b
    \\
   6014.830  &     &   C I   26.07     &    8.64  & -1.710  &   6014.39  &  3.6
 &    22  &   28  &  -17  & 3.1  &  1  &
    \\
   6016.449  &     &   C I   26.06     &    8.64  & -1.820  &   6015.96  &  1.8
 &     9  &   25  &  -20  & 3.2  &  1  &
    \\
             &     &  unident          &          &         &   6017.15  &  2.4
 &     5  &   10  &       & 3.2  &  1  &
    \\
   6046.233  &     &   O I   22        &   10.99  & -1.895  &   6045.77  &  1.5
 &    13  &   42  &  -18  & 3.1  &  1  & b
    \\
   6052.674  &  *  &   S I   10        &    7.87  & -0.740  &   6052.47  &  2.2
 &    11  &   24  &  -18  & 4.2  &  1  &
    \\
   6100.445  &  *  &   C I   37        &    8.85  & -4.520  &   6100.25  &  2.0
 &    12  &   27  &  -17  & 4.2  &  1  &
    \\
   6155.971  &     &   O I   10        &   10.74  & -1.051  &   6155.53  &  5.8
 &    36  &   28  &  -17  & 3.1  &  1  & b
    \\
   6156.778  &     &   O I   10        &   10.74  & -0.731  &   6156.28  &  7.4
 &    45  &   28  &  -20  & 3.1  &  1  & b
    \\
   6158.187  &     &   O I   10        &   10.74  & -0.441  &   6157.70  &  9.8
 &    59  &   28  &  -19  & 3.2  &  1  &
    \\
   6300.311  &     & [OI]    1F        &    0.00  & -9.819  &   6299.57  & -3.9
 &   -11  &   13  &  -31  & 3.2  &  1  & CS, f
    \\
   6453.602  &     &   O I   9         &   10.74  & -1.364  &   6453.12  &  1.2
 &     8  &   27  &  -18  & 3.1  &  1  & b
    \\
   6454.444  &     &   O I   9         &   10.74  & -1.144  &   6453.96  &  1.7
 &    11  &   27  &  -18  & 3.1  &  1  & b
    \\
   6455.977  &     &   O I   9         &   10.74  & -0.994  &   6455.51  &  1.8
 &    15  &   37  &  -17  & 3.1  &  1  &
    \\
   6562.817  &     &   H I   1         &   10.15  &  0.710  &   6561.31  & 61.9
 &  2537  &  103  &  -64  & 3.2  &  3  & H$\alpha$
    \\
   6587.610  &     &   C I   22        &    8.54  & -1.596  &   6587.13  & 10.9
 &    68  &   27  &  -18  & 3.1  &  1  & b
    \\
   6655.517  &     &   C I   21.03     &    8.54  & -1.370  &   6655.05  &  1.2
 &    11  &   41  &  -17  & 3.2  &  1  &
    \\
             &  *  &  unident          &          &         &   6735.37  &  4.1
 &    14  &   14  &       & 2.1  &  1  &
    \\
   6743.531  &     &   S I   8         &    7.87  & -0.920  &   6743.12  &  1.3
 &     8  &   17  &  -14  & 3.2  &  1  &
    \\
   6748.837  &     &   S I   8         &    7.87  & -0.600  &   6748.26  &  2.1
 &    26  &   20  &  -21  & 3.2  &  1  &
    \\
   6757.171  &     &   S I   8         &    7.87  & -0.310  &   6756.69  &  2.5
 &    19  &   30  &  -17  & 3.1  &  1  &
    \\
   6828.115  &     &   C I   21        &    8.54  & -1.280  &   6827.64  &  2.9
 &    24  &   35  &  -17  & 3.2  &  1  &
    \\
   7093.237  &     &   C I   26.01     &    8.65  & -3.260  &   7092.82  &  2.0
 &    13  &   25  &  -13  & 3.1  &  1  &
    \\
   7100.124  &     &   C I   25.02     &    8.64  & -1.600  &   7099.55  &  3.6
 &    24  &   27  &  -20  & 3.1  &  1  &
    \\
   7108.934  &     &   C I   25.02     &    8.64  & -1.680  &   7108.40  &  3.3
 &    18  &   22  &  -18  & 3.1  &  1  &
    \\
   7111.472  &     &   C I   26        &    8.64  & -0.810  &   7110.95  &  6.0
 &    44  &   29  &  -17  & 3.1  &  1  &
    \\
   7113.178  &     &   C I   26        &    8.65  & -0.350  &   7112.63  & 11.3
 &    72  &   25  &  -19  & 3.1  &  2  &
    \\
   7115.172  &     &   C I   25.02     &    8.64  & -0.710  &   7114.66  & 11.3
 &    75  &   26  &  -17  & 3.1  &  1  & b
    \\
%
   7116.991  &     &   C I   25.02     &    8.65  & -0.910  &   7116.44  & 11.6
 &    76  &   26  &  -19  & 3.1  &  2  &
    \\
   7119.656  &     &   C I   25.02     &    8.64  & -1.220  &   7119.11  &  8.3
 &    56  &   27  &  -18  & 3.1  &  1  & b
    \\
   7423.641  &     &   N I   3         &   10.33  & -0.760  &   7423.03  &  5.1
 &    43  &   32  &  -20  & 3.1  &  1  &
    \\
   7442.298  &     &   N I   3         &   10.33  & -0.454  &   7441.70  &  8.9
 &    70  &   30  &  -20  & 3.1  &  1  &
    \\
   7468.312  &     &   N I   3         &   10.34  & -0.270  &   7467.72  & 12.8
 &    92  &   27  &  -19  & 3.1  &  2  &
    \\
   7476.176  &     &   C I   29.03     &    8.77  & -0.760  &   7475.64  &  1.8
 &     9  &   18  &  -17  & 3.1  &  1  &
    \\
   7483.445  &     &   C I   29.03     &    8.77  & -0.490  &   7482.96  &  3.1
 &    22  &   27  &  -15  & 3.1  &  1  &
    \\
   7771.944  &     &   O I   1         &    9.15  &  0.324  &   7771.40  & 55.6
 &   493  &   32  &  -17  & 3.1  &  2  &
    \\
   7774.166  &     &   O I   1         &    9.15  &  0.174  &   7773.61  & 53.5
 &   436  &   30  &  -17  & 3.1  &  3  &
    \\
   7775.388  &     &   O I   1         &    9.15  & -0.046  &   7774.83  & 47.6
 &   377  &   29  &  -17  & 3.1  &  3  &
    \\
             &  *  &  unident          &          &         &   7851.38  &  1.0
 &     6  &   20  &       & 2.1  &  1  &
    \\
   7852.862  &  *  &   C I   32        &    8.85  & -1.420  &   7852.05  &  2.5
 &    14  &   20  &  -17  & 2.1  &  1  &
    \\
   7860.889  &  *  &   C I   32        &    8.85  & -0.730  &   7860.07  &  3.7
 &    31  &   30  &  -17  & 2.1  &  1  &
    \\
             &     &  unident          &          &         &   7946.79  &  2.4
 &    18  &   27  &       & 3.1  &  1  &
    \\
   8058.621  &     &   C I   30.01     &    8.85  & -1.180  &   8058.00  &  4.1
 &    32  &   27  &  -19  & 3.1  &  1  &
    \\
   8184.861  &     &   N I   2         &   10.33  & -0.418  &   8184.26  & 11.2
 &    66  &   20  &  -18  & 3.1  &  1  & b
    \\
   8188.012  &     &   N I   2         &   10.33  & -0.431  &   8187.41  & 10.9
 &    98  &   31  &  -18  & 3.1  &  1  &
    \\
   8223.128  &     &   N I   2         &   10.33  & -0.387  &   8222.50  & 11.5
 &   120  &   36  &  -18  & 3.1  &  1  & b
    \\
   8242.389  &     &   N I   2         &   10.34  & -0.381  &   8241.78  & 10.1
 &   109  &   37  &  -18  & 3.1  &  1  & b
    \\
   8345.553  &     &   H I   11        &   12.04  &         &   8344.86  & 17.5
 &   648  &   96  &  -20  & 3.1  &  1  & P23, b
    \\
   8359.006  &     &   H I   11        &   12.04  &         &   8358.22  & 21.0
 &   759  &  119  &  -24  & 3.1  &  1  & P22, b
    \\
   8374.478  &     &   H I   11        &   12.04  &         &   8373.82  & 23.4
 &  1079  &  125  &  -19  & 3.1  &  1  & P21, b
    \\
   8392.400  &     &   H I   11        &   12.04  & -1.892  &   8391.70  & 28.6
 &  1441  &  115  &  -21  & 3.1  &  1  & P20, b
    \\
   8413.321  &     &   H I   10        &   12.04  & -1.823  &   8412.69  & 33.1
 &  1717  &  126  &  -18  & 3.1  &  1  & P19,b
    \\
   8446.359  &  *  &   O I   14        &    9.52  &  0.170  &   8445.41  & 41.0
 &   325  &   26  &  -19  & 2.1  &  2  & b
    \\
   8446.758  &  *  &   O I   14        &    9.52  & -0.050  &   8445.91  & 39.0
 &   309  &   46  &  -16  & 2.1  &  2  & b
    \\
   8467.256  &     &   H I   10        &   12.04  & -1.670  &   8466.65  & 34.8
 &        &       &  -17  & 3.1  &  1  & P17, half profile, b
    \\
   8502.487  &     &   H I   10        &   12.04  & -1.586  &   8501.83  & 42.4
 &  3029  &  161  &  -19  & 3.1  &  1  & P16, b
    \\
   8542.089  &     &  Ca II  2         &    1.69  &         &   8541.62  &  6.7
 &    34  &   17  &  -12  & 3.1  &  1  & b
    \\
   8545.384  &     &   H I   10        &   12.04  & -1.495  &   8544.71  & 40.2
 &        &       &  -19  & 3.1  &  1  & P15, half profile, b
    \\
   8594.000  &     &   N I   8         &   10.68  & -0.367  &   8593.46  &  4.3
 &    26  &   20  &  -14  & 3.1  &  1  & b
    \\
   8598.394  &     &   H I   9         &   12.04  & -1.398  &   8597.77  & 39.1
 &        &       &  -17  & 3.1  &  1  & P14, half profile, b
    \\
   8629.235  &     &   N I   8         &   10.69  &  0.027  &   8628.53  & 17.8
 &   160  &   29  &  -20  & 3.1  &  2  &
    \\
   8655.878  &     &   N I   8         &   10.69  & -0.650  &   8655.19  &  5.6
 &    57  &   33  &  -20  & 3.1  &  1  &
    \\
   8662.140  &     &  Ca II  2         &    1.69  & -0.730  &   8661.68  &  4.1
 &    26  &   21  &  -11  & 3.1  &  1  & b
    \\
   8665.021  &     &   H I   9         &   12.04  & -1.291  &   8664.35  & 43.8
 &  2383  &  129  &  -19  & 3.1  &  2  & P13, b
    \\
   8680.282  &     &   N I   1         &   10.34  &  0.236  &   8679.58  & 28.7
 &   287  &   32  &  -20  & 3.1  &  2  & b
    \\
   8683.403  &     &   N I   1         &   10.33  & -0.045  &   8682.73  & 22.7
 &   215  &   31  &  -19  & 3.1  &  1  &
    \\
   8728.910  &     &   N I   1         &   10.33  & -1.164  &   8728.19  &  3.6
 &    40  &   36  &  -20  & 3.1  &  1  & b
    \\
   8750.475  &     &   H I   9         &   12.04  & -1.175  &   8749.82  & 46.8
 &  2543  &  115  &  -18  & 3.1  &  3  & P12, b
    \\
   8862.787  &     &   H I   9         &   12.04  & -1.046  &   8862.19  & 43.8
 &        &       &  -16  & 3.1  &  1  & P11, half profile, b
    \\
             &     &  unident          &          &         &   8872.63  &  4.0
 &    35  &   28  &       & 3.1  &  1  & b
    \\
   8873.364  &  *  &   C I   49        &    9.00  & -1.080  &   8873.04  &  4.0
 &    38  &   31  &  -19  & 4.1  &  1  & b
    \\
             &     &  unident          &          &         &   8873.81  &  2.1
 &    19  &   28  &       & 3.1  &  1  & b
    \\
   8874.423  &  *  &   S I   21        &    8.42  & -0.050  &   8874.07  &  3.0
 &    17  &   18  &  -20  & 4.1  &  1  & b
    \\
   9014.911  &     &   H I   9         &   12.04  & -0.901  &   9014.28  & 48.2
 &        &       &  -16  & 3.1  &  1  & P10, half profile, b
    \\
   9061.436  &     &   C I   3         &    7.48  & -0.335  &   9060.87  & 58.0
 &   514  &   28  &  -14  & 3.1  &  1  & b
    \\
   9062.487  &     &   C I   3         &    7.48  & -0.432  &   9061.85  & 58.0
 &   460  &   25  &  -17  & 3.1  &  1  & b
    \\
   9078.288  &     &   C I   3         &    7.48  & -0.557  &   9077.64  & 49.6
 &   437  &   27  &  -17  & 3.1  &  2  & b
    \\
   9088.513  &     &   C I   3         &    7.48  & -0.432  &   9087.85  & 53.7
 &   516  &   30  &  -17  & 3.1  &  1  & b
    \\
   9111.850  &  *  &   C I   13        &    7.49  & -0.335  &   9110.87  & 54.0
 &   556  &   30  &  -18  & 2.1  &  3  &
    \\
   9212.910  &     &   S I   1         &    6.50  &  0.41   &   9212.31  & 38.6
 &   506  &   40  &  -15  & 3.1  &  3  &
    \\
\tabendartvii
\end{table*}

\begin{table*}
\caption{continued}
\tabheaderartvii
   9228.110  &     &   S I   1         &    6.50  &  0.26   &   9227.45  &
 &        &       &       & 3.1  &  1  & blended by H I 9
    \\
   9229.020  &     &   H I   9         &   12.04  & -0.735  &   9228.05  & 49.1
 &        &       &  -27  & 3.1  &  1  & P9, half profile, b
    \\
   9237.490  &     &   S I   1         &    6.50  &  0.04   &   9236.85  & 28.4
 &   340  &   35  &  -16  & 3.1  &  3  &
    \\
             &     &  unident          &          &         &  10016.54  &  9.1
 &    40  &   12  &       & 3.1  &  2  &
    \\
  10049.400  &     &   H I   8         &   12.04  & -0.303  &  10048.83  & 34.0
 &        &       &  -12  & 3.1  &  1  & P$\delta$, half profile, b
    \\
             &     &  unident          &          &         &  10127.23  &  3.2
 &    61  &   53  &       & 3.1  &  1  &
    \\
  10729.530  &     &   C I   1         &    7.48  & -0.401  &  10728.83  & 10.7
 &   109  &   27  &  -15  & 3.1  &  1  &
    \\
\tabendartvii
\end{table*}
\hfill

\end{document}